\documentclass[traditabstract]{aa}
\usepackage{hyperref}
\hypersetup{
    pdftoolbar=true,            % show Acrobat?s toolbar?
    pdfmenubar=true,            % show Acrobat?s menu?
    pdffitwindow=false,         % window fit to page when opened
    pdfstartview={FitH},        % fits the width of the page to the window
    pdftitle={title},                   % title
    pdfauthor={Author},         % author
    pdfsubject={Subject},       % subject of the document
    pdfcreator={Author},        % creator of the document
    pdfproducer={Author},       % producer of the document
    pdfkeywords={AGN}{Stellar Content} , % list of keywords
    pdfnewwindow=true,     % links in new window
    colorlinks=true,            % false: boxed links; true: colored links
    linkcolor=cyan,             % color of internal links (change box color with linkbordercolor)
    citecolor=cyan,             % color of links to bibliography
    filecolor=cyan,                     % color of file links
    urlcolor=cyan               % color of external links
}

\usepackage{natbib}
\usepackage{lscape}
\usepackage{wasysym}
\usepackage{graphicx}
\usepackage{txfonts}
\usepackage{color}

\usepackage[T1]{fontenc}
\newcommand{\PSS}{\textrm{PSS}}
\newcommand{\ESS}{\textrm{ESS}}

\newcommand{\PL}{{\sc PL}}
\newcommand{\Starlight}{{\sc Starlight}}
\newcommand{\Rebetiko}{{\sc Rebetiko}}
\newcommand{\npl}{\texttt{npl}}
\newcommand{\spl}{\texttt{spl}}
\newcommand{\mpl}{\texttt{mpl}}

\begin{document}

% !!!!!!!!!!!!!!!!!!!!!!!!!!!!!!!!!!!!!!!!!!!!!!!!!!!!!!!!!!!!!!!!!!!!!!!!!!!!!!!!!!!!!!!!!!!!!!!!!!!!!!!!!!!!!!!!!!!!!!!!!!!!!!!!!!!!!!!!!!!!!!!!!!!!!!!!!!!!!!!!!!!!!!!!!!!!!!!!!!!!!!!!!!!!!!!
%- - - - - - - - - - - - - - - - - - - - - - - - - - - - - - FRONT MATTER - - - - - - - - - - - - - - - - - - - - - - - - - - - - - - - - - - - - - - -
% !!!!!!!!!!!!!!!!!!!!!!!!!!!!!!!!!!!!!!!!!!!!!!!!!!!!!!!!!!!!!!!!!!!!!!!!!!!!!!!!!!!!!!!!!!!!!!!!!!!!!!!!!!!!!!!!!!!!!!!!!!!!!!!!!!!!!!!!!!!!!!!!!!!!!!!!!!!!!!!!!!!!!!!!!!!!!!!!!!!!!!!!!!!!!!!
\titlerunning{Impact of an AGN featureless continuum on estimation of stellar population properties}
\authorrunning{Cardoso, Gomes \& Papaderos}
\title{Impact of an AGN featureless continuum \\on estimation of stellar population properties}

% --- AUTHORS :::::::::::::::::::::::::::::::::::::::::::::::::::::::::::::::
\author{
Leandro S.~M. Cardoso\inst{\ref{inst1}}$^{\textrm{,}}$\inst{\ref{inst2}}
\and Jean Michel Gomes\inst{\ref{inst1}}
\and Polychronis Papaderos\inst{\ref{inst1}}
}

\institute{
% Institute 1
Instituto de Astrof\' isica e Ci\^ encias do Espa\c co, Universidade do Porto,  CAUP, Rua das Estrelas, PT4150-762 Porto, Portugal\label{inst1}\\
\email{	{\href{mailto:leandro.cardoso@astro.up.pt}{leandro.cardoso@astro.up.pt}},
	 	{\href{mailto:jean@astro.up.pt}{jean@astro.up.pt}}, or
		{\href{mailto:papaderos@astro.up.pt}{papaderos@astro.up.pt}}}
% Institute 2
\and
Departamento de F\'isica e Astronomia, Faculdade de Ci\^encias, Universidade do Porto, 
        Rua do Campo Alegre 687, PT4169-007, Porto, Portugal\label{inst2}
}
% --- AUTHORS :::::::::::::::::::::::::::::::::::::::::::::::::::::::::::::::

% --- DATE ::::::::::::::::::::::::::::::::::::::::::::::::::::::::::::::::::
\date{Received ?? ; Accepted ??}
% --- DATE ::::::::::::::::::::::::::::::::::::::::::::::::::::::::::::::::::

\abstract{
        The effect of the featureless power-law (\PL) continuum of an active galactic nucleus (AGN) on the estimation of physical properties of galaxies with optical population spectral synthesis (\PSS) remains largely unknown. With the goal of a quantitative examination of this issue, we fit synthetic galaxy spectra representing a wide range of galaxy star formation histories (SFHs) and including distinct PL contributions of the form $F_{\nu} \propto \nu^{-\alpha}$ with the \PSS\ code \Starlight\ to study to which extent various inferred quantities (e.g. stellar mass, mean age, and mean metallicity) match the input. The synthetic spectral energy distributions (SEDs) computed with our evolutionary spectral synthesis code include an AGN \PL\ component with $0.5 \leq \alpha \leq 2$ and a fractional contribution $0.2 \leq x_{\mathrm{AGN}} \leq 0.8$ to the monochromatic flux at 4020 \AA. At the empirical AGN detection threshold $x_{\mathrm{AGN}}\simeq 0.26$ that we previously inferred in a pilot study on this subject, our results show that the neglect of a PL component in spectral fitting can lead to an overestimation by $\sim$2 dex in stellar mass and by up to $\sim$1 and $\sim$4 dex in the light- and mass-weighted mean stellar age, respectively, whereas the light- and mass-weighted mean stellar metallicity are underestimated by up to $\sim$0.3 and $\sim$0.6 dex, respectively. These biases, which become more severe with increasing $x_{\mathrm{AGN}}$, are essentially independent of the adopted SFH and show a complex behaviour with evolutionary time and $\alpha$. Other fitting set-ups including either a single PL or multiple PLs in the base reveal, on average, much lower unsystematic uncertainties of the order of those typically found when fitting purely stellar SEDs with stellar templates, however, reaching locally up to $\sim$1, 3 and 0.4 dex in mass, age and metallicity, respectively. Our results underscore the importance of an accurate modelling of the AGN spectral contribution in \PSS\ fits as a minimum requirement for the recovery of the physical and evolutionary properties of stellar populations in active galaxies. In particular, this study draws attention to the fact that the neglect of a PL in spectral modelling of these systems may lead to substantial overestimates in stellar mass and age, thereby leading to potentially significant biases in our understanding of the co-evolution of AGN with their galaxy hosts.
        }

\keywords{galaxies: active -- galaxies: Seyfert -- galaxies: stellar content -- galaxies: evolution}

\maketitle

\markboth{Impact of an AGN featureless continuum on estimation of stellar population properties}{Impact of an AGN featureless continuum on estimation of stellar population properties}

% !!!!!!!!!!!!!!!!!!!!!!!!!!!!!!!!!!!!!!!!!!!!!!!!!!!!!!!!!!!!!!!!!!!!!!!!!!!!!!!!!!!!!!!!!!!!!!!!!!!!!!!!!!!!!!!!!!!!!!!!!!!!!!!!!!!!!!!!!!!!!!!!!!!!!!!!!!!!!!!!!!!!!!!!!!!!!!!!!!!!!!!!!!!!!!!
% - - - - - - - - - - - - - - - - -- - - - - - - - - - - -  - - INTRODUCTION   - - - - - - - - - - - - - - - - - - - - - - - - - - - - - - - - - - - - -
% !!!!!!!!!!!!!!!!!!!!!!!!!!!!!!!!!!!!!!!!!!!!!!!!!!!!!!!!!!!!!!!!!!!!!!!!!!!!!!!!!!!!!!!!!!!!!!!!!!!!!!!!!!!!!!!!!!!!!!!!!!!!!!!!!!!!!!!!!!!!!!!!!!!!!!!!!!!!!!!!!!!!!!!!!!!!!!!!!!!!!!!!!!!!!!!
\section{Introduction}\label{Sec:Introduction}

        The mass assembly, star formation and chemical evolution history of galaxies are fundamental to the understanding of the physical and spectrophotometric properties of galaxies in the local universe. In recent decades, numerous studies have employed spectral synthesis with the aim of addressing these issues. This technique saw its initial realisation  with the works by \cite{Whipple_1935} and \cite{Baade_1944a, Baade_1944b}. They have attempted to categorise groups of stars within local galaxies according to their spectral types and found that, predominantly, O and B stars were in the disk and G, K, and M stars were in the bulge. This was later recognised to be a signature of the different stellar populations (population I, II, and III stars) composing distinct structural components in galaxies. These works planted the seed of the modern concept for modelling the spectral continuum of galaxies as due to the linear superposition of individual star spectra or star clusters in galaxies. Historically, spectral synthesis has followed two main approaches: population spectral synthesis (\PSS) and evolutionary spectral synthesis (\ESS).

        On the one hand, also known as population synthesis or inversion technique, PSS aims at decomposing the observed spectrum of a galaxy into its main elementary building blocks, such as individual stars and/or groups of stars of a given age, metallicity, and stellar initial mass function (IMF). Per definition, this technique yields a discretised approximation to the star formation and chemical evolution history  (SFH and CEH, respectively) of a galaxy. The \PSS\ approach saw its inception with the works by \cite{Morgan_1956}, \cite{Wood_1966}, and \cite{Faber_1972} and has over the years been subject to significant development (e.g. \citealt{Bica_1988, Pelat_1997, Pelat_1998, Moultaka_etal_2004, Heavens_Jimenez_Lahav_2000, Heavens_etal_2004, CidFernandes_etal_2005, Ocvirk_etal_2006a, Ocvirk_etal_2006b, Tojeiro_etal_2007, MarcArthur_etal_2009, Koleva_etal_2009}). In the last two decades, spectral synthesis has transitioned from the modelling of a few observables, such as colours, fluxes, and/or equivalent widths (EWs) of absorption lines (e.g. \citealt{Wood_1966, Faber_1972, Worthey_1994, Kauffmann_etal_2003c}) to the more powerful pixel-by-pixel ($\lambda$-by-$\lambda$) fitting technique that exploits the full information in current medium- to high-resolution  spectroscopic data (e.g. \citealt{CidFernandes_etal_2005, Koleva_etal_2009, Tojeiro_etal_2007}). 

        On the other hand, also known as the direct approach, \ESS\ allows for the estimation of physical properties of galaxies, such as the stellar mass, age, and metallicity, on the basis of empirically founded assumptions on the IMF, chemical evolution, and star formation rate (SFR). This approach was pioneered by \cite{Tinsley_1968} and \cite{Spinrad_Taylor_1972} and since then has been  a thriving topic of research with the creation of increasingly sophisticated models (e.g. \citealt{Kruger_etal_1995, Fioc_RoccaVolmerange_1997, Leitherer_etal_1999, Zackrisson_etal_2001, Bruzual_Charlot_2003, Anders_etal_2004, LeBorgne_etal_2004, Maraston_2005, Coelho_etal_2007, Molla_etal_2009, Kotulla_etal_2009, Vazdekis_etal_2010}).

        Whereas there is broad consensus that these techniques have led in the past decade to a great leap forward  in our understanding of the assembly history of normal galaxies, their application to active galactic nucleus (AGN) remains controversial and problematic. As a matter of fact, the chief goal of spectral modelling of AGN spectra in the   70s and 80s has been the recovery of the AGN emission after fitting and removal of the underlying stellar  spectral energy distribution (SED). The main assumption behind this approach was that the underlying host galaxy was composed solely of old and metal-rich stellar populations that are in general well represented by an elliptical galaxy spectral template.  For instance, the original work by \citet{Koski_1978} showed, using photometric scans of a sample of Seyferts (named after Carl Keenan Seyfert; \citealt{Seyfert_1943}) and narrow-line radio galaxies, that a non-stellar component could be  approximated by a featureless continuum (FC) with a power-law (\PL) function $F_{\nu} \propto \nu^{-\alpha}$ in the UV-optical range with $\alpha$ between 0.2 and 2.8 and an average value of $\langle \alpha \rangle = 1.0\pm0.5$. In this study, the AGN was found to provide on average $\sim$30\% of the continuum at the H$\beta$ wavelength.

        The accuracy of these methods for removing the underlying stellar SED has generally been limited by various sources of uncertainty (as extensively discussed in \citealt{Ho_Filippenko_Sargent_1997}, and more recently in \citealt{Ho_2008}) and, from the spectral synthesis point of view, the techniques employed were not meant to investigate, or were not capable of exploring, the SFH and CEH of galaxies. As a consequence, it might be questioned whether these studies could accurately retrieve the pure SED of the AGN, in particular for galaxies where stellar emission dominates. Even though more elaborate methods were developed for the starlight removal, such as off-nuclear spectrum subtraction within the same galaxy  \citep{StochiBergmann_etal_1993} or a linear combination of the spectra of  different galaxies \citep{Ho_Filippenko_Sargent_1997,Ho_Filippenko_Sargent_2003, StochiBergmann_etal_1998}, they generally did not  have the reconstruction of the stellar mass assembly history of active galaxies as prime objective.
        
        Concerning the \PL\ contribution, it is well-established in the framework of the AGN unified model (\citealt{Antonucci1993, Netzer2015}) that it constitutes the dominant component in the SED optical continuum  of type~1 AGNs, such as quasars and Seyfert 1 galaxies.  However, its relative contribution to the UV-optical in type~2 AGN, such as Seyfert~2 and low-ionisation nuclear emission-line region (LINER) galaxies (\citealt{Heckman_1980}), remained controversial, largely  owing to the limited capability of current spectral fitting codes.

     Indeed, even though a substantial body of work has been devoted to this subject, a review of the literature reveals  a wide set of discordant conclusions, in some cases drawn using similar methods. Part of the confusion might be attributed to poorly described or justified methodological assumptions and the assessment  of the propagation of uncertainties in the physical properties  of both the estimated stellar and AGN contribution,  besides sample selection effects.

        For example, observations of the optical stellar features, such as the Mgb band ($\sim$5200\AA), have shown in some cases a certain degree of dilution in these regions by a \PL\ component (e.g. \citealt{Koski_1978, Dressler1984, NelsonWhittle1995, SeroteRoos_etal_1998, CidFernandes1998, Boisson_etal_2000, Moultaka_Pelat_2000,  Kauffmann_etal_2003c, GarciaRissmann_etal_2005, Vega_etal_2009}).  Conversely, it was pointed out that the EWs of the infrared Ca II triplet  (hereafter CaT at 8498, 8542, and 8662 \AA) were similar or even larger  than in normal galaxies (e.g. \citealt{Terlevich_Diaz_Terlevich_1990,NelsonWhittle1995}),  as they are more suitable for a stellar population and kinematical analysis. \cite{Vega_etal_2009} applied the \PSS\ code \Starlight\ to active galaxies and found that Seyfert 1 galaxies have EW(CaT) between $\sim$1.5 and $\sim$7.5 \AA\ with a median value of 4.8 \AA\ and that Seyfert~2 lie between $\sim$1.5 and $\sim$10.5 \AA\ with a median value of 6.5 \AA. Their results suggest that Seyfert 2 show no sign of dilution of EW(CaT) as compared to normal galaxies, even though optical absorption lines such as the CaII K (CaK) band at 3933 \AA\ are much weaker than in old, bulge-like stellar  populations.  Evolutionary models in an EW(CaT)--EW(CaK) diagram suggest that young stellar populations are responsible for the dilution of optical lines in active galaxies.  However, the authors concluded that  non-stellar contributions can reach $\sim$85\% in Seyfert 1, $\sim$50 \% in Seyfert~2 and starburst galaxies,  and $\sim$32\% even for normal galaxies by applying \Starlight\ with multiple PLs with $\alpha=1$--2.

        The controversy on the \PL\ contribution to type 2 objects remains until today. For example, in studies by \cite{Ho_Filippenko_Sargent_1995, Ho_Filippenko_Sargent_1997, Ho_Filippenko_Sargent_2003} and \cite{Ho_2008} it was  argued that samples analysed by some other groups were, in general, not large enough and unbiased, preventing an accurate separation of nuclear from global properties. Even though other surveys, such as SDSS (\citealt{York_etal_2000}), have surpassed statistically the number of galaxies of smaller surveys (e.g. the Palomar sample; \citealt{Ho_Filippenko_Sargent_1995}), these larger surveys have the disadvantage  of large projected aperture sizes, where the continuum emission in type 2 AGN could be significantly  diluted by star formation (\citealt{Kauffmann_etal_2003c}). This might explain why some works have found almost no PL contribution in type 2 AGN (\citealt{CidFernandes1998, Schmitt_StorchiBergmann_CidFernandes_1999}, while others found that a PL could account for up $\sim$90\% of the optical flux (e.g. \citealt{Koski_1978, Vega_etal_2009}).

        For instance, \cite{CidFernandes1998} have studied the stellar content of active galaxies with long-slit spectroscopy of 38 active and 4 normal galaxies with the goal of detecting the fractional contribution of a FC relative to the stellar emission. This study detected dilution of stellar absorption lines by a \PL\ in most of the galaxies with broad-line emission, whereas almost no \PL\ contribution was found for most of the type~2 Seyferts in their sample.  Likewise, a follow-up study in  \citet{Schmitt_StorchiBergmann_CidFernandes_1999}  investigated the contribution by an AGN \PL\ to the optical using flux continuum ratios and absorption-line EWs on spectra of nuclear regions of 20 local Seyfert 2 galaxies. The conclusions drawn was that the contribution from stars younger than 10 Myr and a \PL\ rarely exceeds 5\% and that the high FC contribution (up to $\sim$70\%)  found by \citet{Koski_1978} was likely overestimated  because of the use of an elliptical template spectrum for the nuclear stellar population. 

        More recently, \citet{CidFernandes_etal_2004} applied a $\lambda$-by-$\lambda$ spectral synthesis approach with an early version of the \PSS\ code \Starlight\ (\citealt{CidFernandes_etal_2005}) to constrain the SFH of 79 Seyfert~2 galaxies in the 3500--5200 \AA\ range while simultaneously fitting an additional \PL\ continuum with varying spectral index $\alpha$. The authors found stellar populations of all ages and a \PL\  mean contribution of $\sim$20--30\% and maximum contribution of $\sim$63\% to the monochromatic  flux at 4020 \AA. Although the authors warned that the \PL\ component might be due to scattered  light from a hidden AGN or a young and dusty starburst, its importance was clearly  documented in this \PSS\ modelling study. Likewise, \cite{Benitez_etal_2013} found that a \PL\ component provides up to $\sim$30\% to the monochromatic luminosity at 5100 \AA\ by applying \Starlight\ on 10 nearby intermediate-type AGN from SDSS.

        Some contradictory results were also found by \cite{Eracleous_Halpern_2001}, who characterised NGC~3065 as a LINER with broad Balmer emission lines coming  from an accretion disk. The authors modelled the continuum of that galaxy with  a linear combination of a PL non-stellar component with starlight described by different spectra of elliptical S0 galaxies.  This work found that the SED continuum of NGC~3065 can be modelled without the need for a \PL\ if the spectrum of NGC~4339 is taken  as a template for subtracting the stellar contribution, whereas other templates  would imply a non-stellar component contributing up to $\sim$10\% of the SED continuum at H$\alpha$ and H$\beta$ and $\la$15\% at [O{\sc ii}]$\lambda$3727.

        As many of the results obtained by fitting a mixture of stellar and \PL\ templates to {observed} spectra remain divergent and the estimation of stellar population properties with this technique is  still largely uncharted territory\footnote{Whether a second and dominant blue featureless continuum (FC2) is present in type 2 AGNs due to compact starbursts (e.g. \citealt{Cid1995,Heckman1995,Tran1995}), instead of coming only from the AGN synchrotron emission, is irrelevant to our work, which focusses on the recovery of the stellar population properties (e.g. SFHs and CEHs) of the host galaxy in case a \PL\ (or AGN-like) contribution is present.}, it appears to be worthwhile to supplement existing work with modelling of synthetic  composite spectra with stellar and AGN components of known constitution for the sake of evaluating the capability of  SED fitting codes to retrieve the input physical and evolutionary characteristics of a galaxy (e.g. SFH, CEH, and AGN contribution).  The very few works existing in this regard (e.g. \citealt{Bon_Popovic_Bon_2014,Hayward_Smith_2015, Ciesla_etal_2015,Cardoso_Gomes_Papaderos_2016}) yield a variety of conclusions.

        \cite{Bon_Popovic_Bon_2014} simulated a mock galaxy sample of 7000 integrated spectra of Seyfert 2 galaxies and used the \PSS\ code ULySS (\citealt{Koleva_etal_2009}) to recover the underlying stellar contribution and \PL\ component. Their main results show that the stellar populations characteristics can be retrieved whenever the signal-to-noise ratio is higher than 20 and if the stellar contribution is more than 10\% of the total flux.

        A different methodology was adopted by \cite{Hayward_Smith_2015}, where hydrodynamical simulations with full radiative transfer modelling were used to create mock SEDs with known observational and physical characteristics (e.g. photometric bands, V-band extinction, stellar mass, dust luminosity and mass, and SFR). The authors in turn applied the \textsc{magphys} code (\citealt{daCunha_etal_2008}) with ultraviolet to millimetre photometry in an attempt to recover the stellar population properties. Overall, results showed that most physical parameters are well estimated, although it was found that in certain cases (e.g. major mergers)  AGN contamination leads to a stellar mass overestimation of $\sim$0.03 dex at AGN activity peaks. 

        At variance to the previous studies, a strong dependence on the estimated stellar properties on the specifics of SED fitting  was reported by \cite{Ciesla_etal_2015}. These authors fitted broadband photometric SEDs with \textsc{cigale} (\citealt{Noll_etal_2009}) assuming exponentially decreasing and delayed SFHs, solar-metallicity \cite{Maraston_2005} SSPs, \cite{Calzetti_etal_2000} dust extinction law, and dust remission. Their results showed that exclusion of AGN optical  templates from the fit can lead to the overestimation of the stellar mass by up to $\sim$150\% and of the SFR by up to $\sim$300\%, depending on AGN spectral type and SFH. Both overestimations increase with the AGN overall contribution to the total infrared luminosity $x^{\mathrm{IR}}_{\mathrm{AGN}}$. Inclusion of AGN templates in the fit can lead to the overestimation of the AGN light fraction up to $\sim$100\%, depending on the AGN spectral shape and $x^{\mathrm{IR}}_{\mathrm{AGN}}$. However, the estimated AGN light fraction tends roughly to its true value with increasing $x^{\mathrm{IR}}_{\mathrm{AGN}}$. Quite importantly, the authors found uncertainties in mass up to $\sim$40\% ($\sim$0.15 dex) and $\sim$40--50\% on the SFR, depending on the AGN spectral shape and $x^{\mathrm{IR}}_{\mathrm{AGN}}$.

        Our work builds upon and extends a pilot investigation of biases in optical \PSS\ modelling of early-type active galaxies showing a high Lyman continuum (LyC) photon escape fraction presented in \citet{Cardoso_Gomes_Papaderos_2016} (hereafter CGP16). The observational motivation behind this study comes from the estimation of a high ($>$90\%) LyC photon escape fraction in early-type active galaxies by \citet{Papaderos_etal_2013} and \citet{Gomes_etal_2016c}, which, according to our current knowledge, host a super-massive black hole in their nuclei with expected accretion-powered activity. More generally, it is conceivable that the model framework  adopted in this study is applicable to any other galaxy spheroid or spheroidal component (e.g. classical bulges),  where, in the absence of absorbing gas of sufficient density, the bulk of the LyC radiation from the AGN accretion disk may  escape without producing in situ optical nebular emission. Motivated by such considerations, the main objective of this study is to shed further light into the assembly history of active galaxies by quantifying potential uncertainties in the recovery of their SFHs and CEHs and characteristic stellar population properties with state-of-the-art \PSS. 

        This paper is organised as follows. Section \ref{Sec:Methodology} presents our library of synthetic spectra with stellar and AGN components (Subsection~\ref{SubSec:ESS}) and the adopted PSS modelling approach (Subsection \ref{SubSec:PSS}). Sections \ref{Sec:Results} and \ref{Sec:Discussion} provide an overview and discussion
of our results. Finally, Section \ref{Sec:Conclusions} summarises the main findings and conclusions of this work.

% !!!!!!!!!!!!!!!!!!!!!!!!!!!!!!!!!!!!!!!!!!!!!!!!!!!!!!!!!!!!!!!!!!!!!!!!!!!!!!!!!!!!!!!!!!!!!!!!!!!!!!!!!!!!!!!!!!!!!!!!!!!!!!!!!!!!!!!!!!!!!!!!!!!!!!!!!!!!!!!!!!!!!!!!!!!!!!!!!!!!!!!!!!!!!!!
% - - - - - - - - - - - - - - - - -- - - - - - - - - - - -  - - - -  METHOD   - - - - - - - - - - - - - - - - - - - - - - - - - - - - - - - - - - - - - - - - 
% !!!!!!!!!!!!!!!!!!!!!!!!!!!!!!!!!!!!!!!!!!!!!!!!!!!!!!!!!!!!!!!!!!!!!!!!!!!!!!!!!!!!!!!!!!!!!!!!!!!!!!!!!!!!!!!!!!!!!!!!!!!!!!!!!!!!!!!!!!!!!!!!!!!!!!!!!!!!!!!!!!!!!!!!!!!!!!!!!!!!!!!!!!!!!!!
\section{Methodology}\label{Sec:Methodology}

% !!!!!!!!!!!!!!!!!!!!!!!!!!!!!!!!!!!!!!!!!!!!!!!!!!!!!!!!!!!!!!!!!!!!!!!!!!!!!!!!!!!!!!!!!!!!!!!!!!!!!!!!!!!!!!!!!!!!!!!!!!!!!!!!!!!!!!!!!!!!!!!!!!!!!!!!!!!!!!!!!!!!!!!!!!!!!!!!!!!!!!!!!!!!!!!
\subsection{Synthetic spectra library}\label{SubSec:ESS}
% !!!!!!!!!!!!!!!!!!!!!!!!!!!!!!!!!!!!!!!!!!!!!!!!!!!!!!!!!!!!!!!!!!!!!!!!!!!!!!!!!!!!!!!!!!!!!!!!!!!!!!!!!!!!!!!!!!!!!!!!!!!!!!!!!!!!!!!!!!!!!!!!!!!!!!!!!!!!!!!!!!!!!!!!!!!!!!!!!!!!!!!!!!!!!!!

        A library of synthetic SEDs was computed with the ESS code \Rebetiko\ (Papaderos \& Gomes, in prep.; hereafter PG). These SEDs are a time- and metallicity-dependent linear combination of coeval and chemically homogeneously groups of stars known as simple stellar populations (SSPs), which is a term introduced by \cite{Renzini_1981}. The flux of a composite stellar population (CSP) $F^{\mathrm{CSP}}_\lambda (t)$ representing the overall stellar content of a galaxy of age $t$ can be computed by linearly combining SSPs of different ages and metallicities according to an assumed SFR $\Psi(t) = \mathrm{d}M_{\star}/\mathrm{d}t$,

        \begin{equation}\label{Equation-ESS_CSP}
        F^{\textrm{CSP}}(\lambda,t,Z) =  
                        \int_{0}^{t} \Psi(t-t^{\prime})F^{\textrm{SSP}}(\lambda,t^{\prime},Z) \textrm{ d} t^{\prime}  ,
                                                                                                                                                                \end{equation}

        \noindent where $F^{\textrm{SSP}}(\lambda,t^{\prime},Z)$ represents the SSP spectral library as a function of the wavelength $\lambda$, age $t^{\prime}$, and metallicity $Z$ (see e.g. \citealt{Tinsley_1980, Walcher_etal_2011, Cervino_2013, Conroy_2013} for reviews on spectral synthesis).

%- - - - - - - - - - - - - - - - - - - - - - - - - - - - - - - - - - - - - - - - - - - - - - - - - - - - - - - - - - - - - - - - - - - - - - - - - - - - - - - - - - - - - - - - - - - - 
\begin{figure}
\begin{center}
\includegraphics[width=\columnwidth]{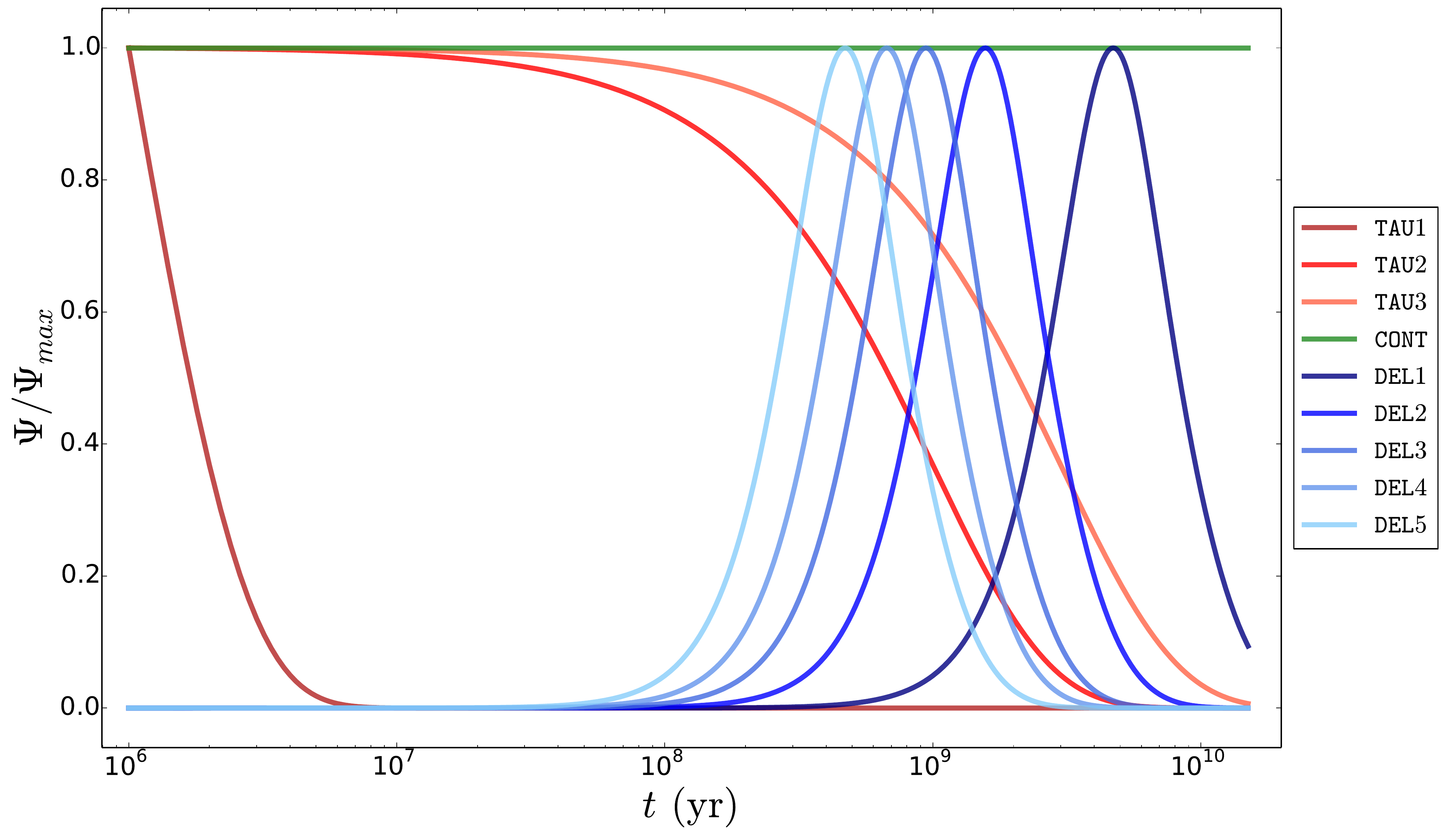}
\caption{SFR functions $\Psi(t)$ adopted in the ESS code \Rebetiko\ for the assembly of purely-stellar evolutionary models. Red, green, and blue lines represent exponentially declining, continuous, and delayed SFR functions, respectively. }
\label{Fig:SFRs_of_REBETIKO}
\end{center}
\end{figure}
%- - - - - - - - - - - - - - - - - - - - - - - - - - - - - - - - - - - - - - - - - - - - - - - - - - - - - - - - - - - - - - - - - - - - - - - - - - - - - - - - - - - - - - - - - - - - 

        This work adopts \cite{Bruzual_Charlot_2003} solar-metallicity ($Z_{\odot}=0.02$) SSPs with a \cite{Chabrier_2003} IMF to create synthetic CSP spectra for a  fixed metallicity and a wide range of SFH parameterisations that are  generally assumed to approximate those of the different Hubble  types (e.g. \citealt{RoccaVolmerange_Guiderdoni_1988, Gavazzi_etal_2002, Bruzual_Charlot_2003}).  Figure \ref{Fig:SFRs_of_REBETIKO} shows the adopted SFR functions: exponentially declining (red lines) with e-folding timescales of 0.001, 1, and 3 Gyr from left to right, respectively;  continuous (green line), and delayed (blue lines) with different star formation peaks as a function of the look-back time (e.g. \citealt{Gavazzi_etal_2002}; PG).
        
        The library comprises 716 spectra with ages  between 1 Myr and 15 Gyr for each SFH complemented with a wide range  of physical properties, such as SFHs weighted by light and mass and CEHs with constant solar metallicity. These purely stellar models are used in this work to provide baseline results to which active galaxy models are compared to when applying PSS.
                
        Another library of more complex models was created by adding simple AGN continua models to the CSPs. The optical AGN FC is assumed to be well represented by a PL defined as $F_{\nu} \propto \nu^{-\alpha}$  (e.g. \citealt{Oke_Neugebauer_Becklin_1970, OConnell_1976,  Koski_1978}). This approximation has been widely adopted in  spectral synthesis (e.g. \citealt{Goerdt_Kollatschny_1998,  Schmitt_StorchiBergmann_CidFernandes_1999, Kauffmann_etal_2003c,  CidFernandes_etal_2004, Moultaka_2005}) and photoionisation studies  (e.g. \citealt{Ferland_Netzer_1983, Stasinska_1984a, Stasinska_1984b, Mathews_Ferland_1987, Veilleux_Osterbrock_1987}), in which the PL index is commonly within $\alpha=0.5$--2.
        
        The flux $F_{\lambda}$ normalised at wavelength $\lambda_0$ is  the linear combination of purely stellar model $F^{\star}_{\lambda}$ and AGN $F^{\mathrm{AGN}}_{\lambda}$ continua, 
        
        \begin{equation}\label{Eq:Linear_Combination}
                F_{\lambda} = x_{\star} \frac{F^{\star}_{\lambda}}{F^{\star}_{\lambda_0}} +
                x_{\mathrm{AGN}} \frac{F^{\mathrm{AGN}}_{\lambda}}{F^{\mathrm{AGN}}_{\lambda_0}} ,
                                                                                                                                                        \end{equation}

        \noindent where $x_{\star}$ and $x_{\mathrm{AGN}}$ are the fractional contributions of stellar and AGN monochromatic fluxes at $\lambda_0=4020$ \AA, respectively. Equation \ref{Eq:Linear_Combination} is constrained by the normalisation condition $x_{\star}+x_{\mathrm{AGN}} = 1$ at $\lambda_0$.

%- - - - - - - - - - - - - - - - - - - - - - - - - - - - - - - - - - - - - - - - - - - - - - - - - - - - - - - - - - - - - - - - - - - - - - - - - - - - - - - - - - - - - - - - - - - - 
\begin{figure}
\begin{center}
\includegraphics[width=\columnwidth]{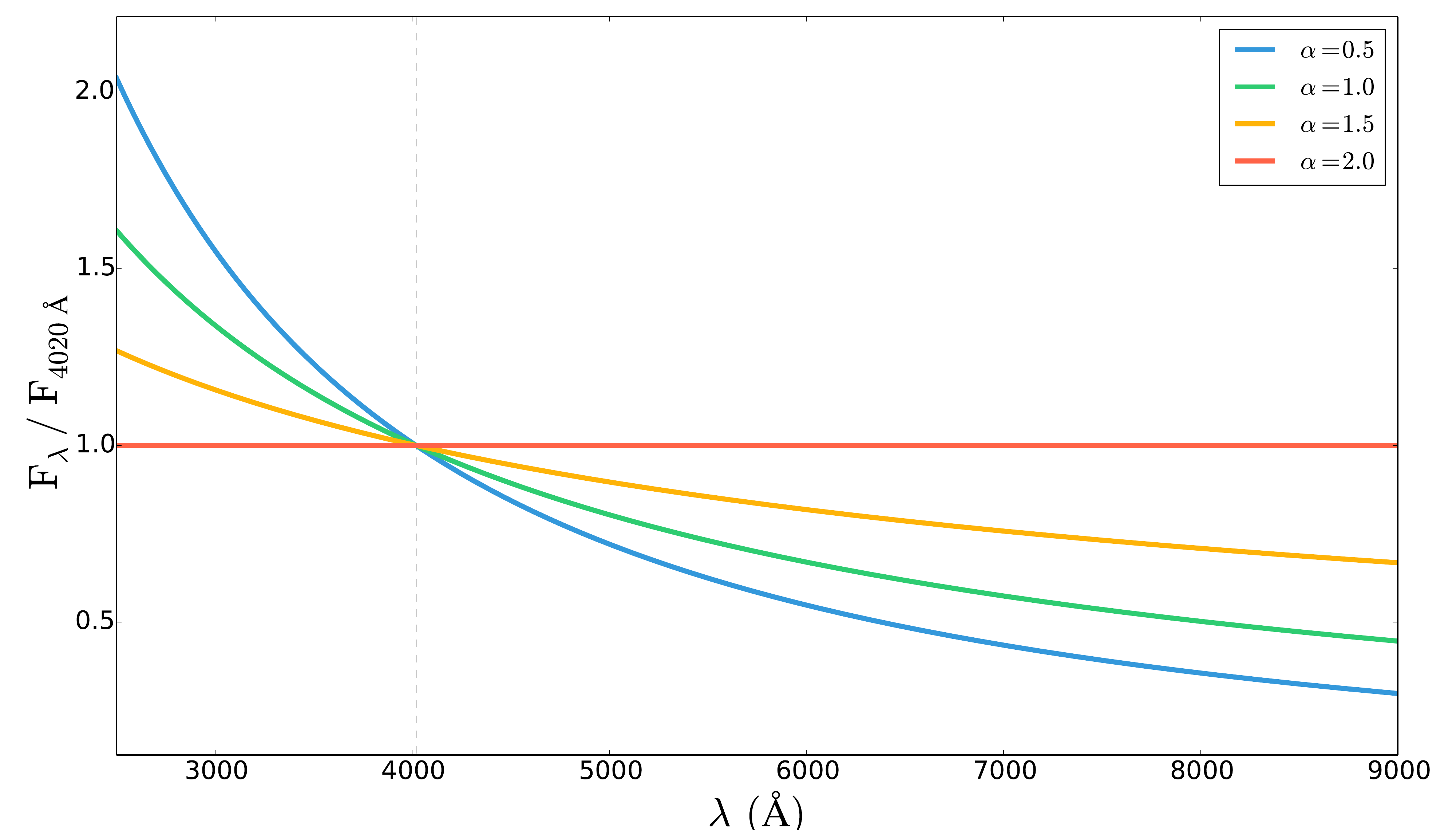}
\caption{Featureless PL AGN continua parameterised as $F_{\nu} \propto \nu^{-\alpha}$ normalised at $\lambda_0=4020$ \AA\ (dashed vertical line) as a function of wavelength $\lambda$. Blue, green, yellow, and red lines represent PLs with $\alpha=0.5$, 1, 1.5, and 2, respectively.}
\label{Fig:AGN_PowerLaws}
\end{center}
\end{figure}
%- - - - - - - - - - - - - - - - - - - - - - - - - - - - - - - - - - - - - - - - - - - - - - - - - - - - - - - - - - - - - - - - - - - - - - - - - - - - - - - - - - - - - - - - - - - - 
                
        It is useful to rewrite Equation \ref{Eq:Linear_Combination} as
                
                \begin{equation}\label{Eq:Linear_Combination_Modified}
                        F_{\lambda} = F^{\star}_{\lambda} + \frac{x_{\mathrm{AGN}}}{x_{\star}}  
                                               F^{\star}_{\lambda_0} \left( \frac{\lambda}{\lambda_0} \right)^{\alpha-2} ,
                                                                                                                                                                \end{equation}
        
        \noindent where the total stellar flux $F^{\star}_{\lambda}$ is unchanged and it is assumed that $F^{\mathrm{AGN}}_{\lambda}\propto \lambda^{\alpha-2}$, following $F^{\mathrm{AGN}}_{\nu} \propto \nu^{-\alpha}$. We adopted AGN PL continua with $\alpha=0.5$, 1, 1.5 and 2 and $x_{\mathrm{AGN}}=0.2$, 0.4, 0.6 and 0.8 to study the effects of AGN PL index and fractional contribution variations in the estimation of stellar population properties with PSS. These wide ranges assure us that no strong prior assumptions are made concerning the AGN optical shape and relative contribution. Figure \ref{Fig:AGN_PowerLaws} shows the adopted AGN PL continua with $\alpha=0.5$, 1, 1.5, and 2 corresponding to blue, green, yellow, and red lines, respectively. This figure shows that the AGN PL becomes bluer with decreasing $\alpha$. 
                
%- - - - - - - - - - - - - - - - - - - - - - - - - - - - - - - - - - - - - - - - - - - - - - - - - - - - - - - - - - - - - - - - - - - - - - - - - - - - - - - - - - - - - - - - - - - - - 
\begin{figure*}[t]
\begin{center}
\includegraphics[width=0.95\textwidth]{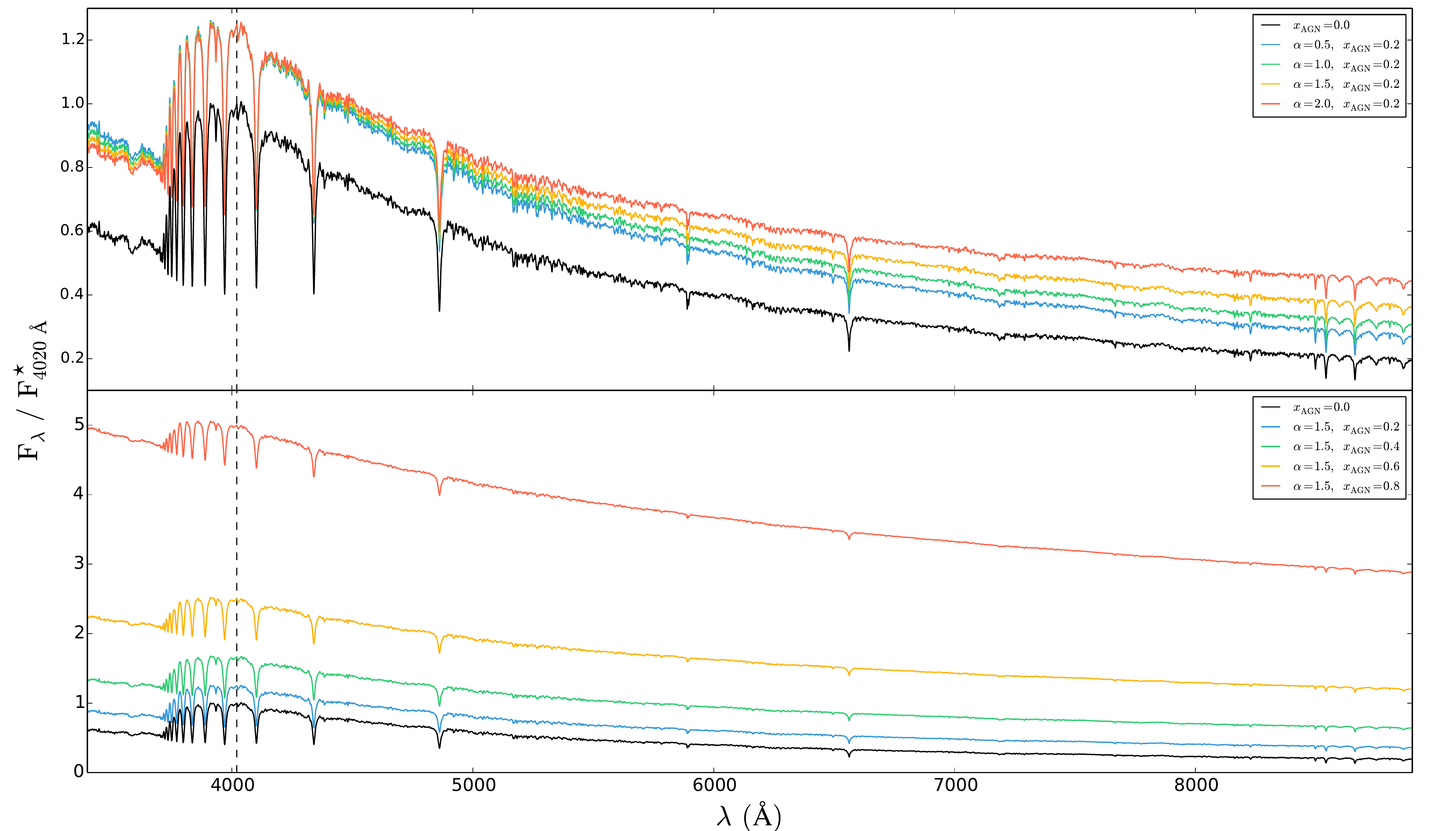}
\caption{Combination of a CSP SED for an instantaneously formed stellar population of age 100 Myr and solar metallicity with AGN PLs for different values of $\alpha$ and fractional contribution $x_{\mathrm{AGN}}$ to the monochromatic flux at $\lambda_0=4020$ \AA\ (black dashed line). Black lines represent purely stellar SEDs and blue, green, yellow, and red lines represent active galaxy SEDs with $\alpha=0.5$, 1, 1.5, and 2 for $x_{\mathrm{AGN}}=0.4$ (top panel) and $x_{\mathrm{AGN}}=0.2$, 0.4, 0.6, and 0.8 for $\alpha=1.5$ (bottom panel), respectively.}
\label{Fig:Linear_Combination_Example}
\end{center}
\end{figure*}
%- - - - - - - - - - - - - - - - - - - - - - - - - - - - - - - - - - - - - - - - - - - - - - - - - - - - - - - - - - - - - - - - - - - - - - - - - - - - - - - - - - - - - - - - - - - - - 

        As an example, Fig. \ref{Fig:Linear_Combination_Example} shows SEDs resulting from combining an instantaneously formed stellar population with 100 Myr and solar metallicity with AGN PLs with $\alpha$ and $x_{\mathrm{AGN}}$ variations. Purely stellar spectra are represented by black lines and models with AGN PLs with $\alpha=0.5$, 1, 1.5, and 2 for $x_{\mathrm{AGN}}=0.4$ (top panel) and $x_{\mathrm{AGN}}=0.2$, 0.4, 0.6, and 0.8 for $\alpha=1.5$ (bottom panel) are represented by blue, green, yellow, and red lines, respectively.    
        
        This figure illustrates two facts of special relevance to this study. Firstly, the normalisation wavelength $\lambda_0$ defines how the AGN continuum shape affects the underlying stellar continuum since Eq. \ref{Eq:Linear_Combination_Modified} implies a flux increase  when $\alpha<2$ larger than $F_{\star,\lambda_0}$ for $\lambda<\lambda_0$ and smaller than $F_{\star,\lambda_0}$ for $\lambda>\lambda_0$. Thus,  adopting typical values of $\lambda_0$ such as 4020 \AA\ (e.g. \citealt{CidFernandes_etal_2004}) or 5050 \AA\ (e.g. \citealt{Goerdt_Kollatschny_1998}) means that the AGN  continuum approximated as a PL for $\alpha<2$ has a slope similar to that of young SSPs. Secondly, the dilution effect of absorption-line features due to the addition of a FC (e.g. \citealt{Koski_1978, SeroteRoos_etal_1998, Moultaka_Pelat_2000, Kauffmann_etal_2003c}) depends on adopted $\lambda_0$, $\alpha,$ and $x_{\mathrm{AGN}}$.

\begin{figure}
\begin{center}
\includegraphics[width=\columnwidth]{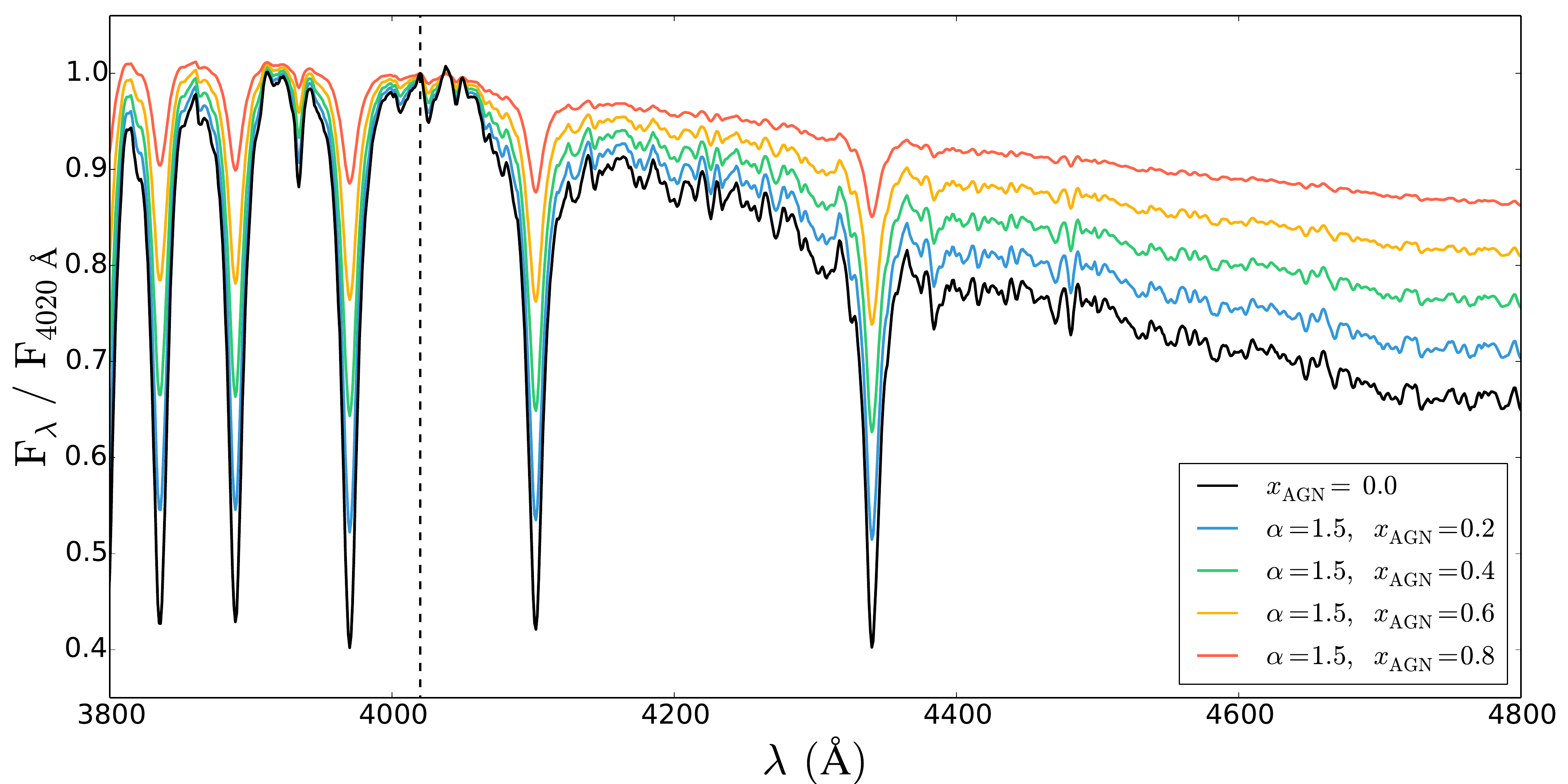}
\caption{Synthetic SEDs normalised at $\lambda_0$ (black dashed line) with the contributions of an instantaneously formed solar-metallicity stellar population of age 100 Myr and an AGN PL with fixed $\alpha=1.5$. The black line represents the purely stellar continuum and the blue, green, yellow, and red lines represent active galaxy SEDs with $x_{\mathrm{AGN}}=0.2$, 0.4, 0.6, and 0.8, respectively. }
\label{Fig:Dilution}
\end{center}
\end{figure}

        Figure \ref{Fig:Dilution} illustrates this last point by presenting SEDs normalised at $\lambda_0$ for an instantaneous burst SFH  with 100 Myr and solar metallicity (black line) combined with AGN continua for fixed $\alpha=1.5$ and $x_{\mathrm{AGN}}=0.2$, 0.4, 0.6, and 0.8 (blue, green, yellow, and red lines, respectively). This figure shows that $\lambda_0$ defines a demarcation point in the spectra that separates regions suffering from blueing or reddening. Moreover, the dilution of the underlying stellar continuum induced by the AGN PL increases with increasing $x_{\mathrm{AGN}}$.
                
%- - - - - - - - - - - - - - - - - - - - - - - - - - - - - - - - - - - - - - - - - - - - - - - - - - - - - - - - - - - - - - - - - - - - - - - - - - - - - - - - - - - - - - - - - - - - - 
%- - - - - - - - - - - - - - - - - - - - - - - - - - - - - - - - - - - - - - - - - - - - - - - - - - - - - - - - - - - - - - - - - - - - - - - - - - - - - - - - - - - - - - - - - - - - - 

% !!!!!!!!!!!!!!!!!!!!!!!!!!!!!!!!!!!!!!!!!!!!!!!!!!!!!!!!!!!!!!!!!!!!!!!!!!!!!!!!!!!!!!!!!!!!!!!!!!!!!!!!!!!!!!!!!!!!!!!!!!!!!!!!!!!!!!!!!!!!!!!!!!!!!!!!!!!!!!!!!!!!!!!!!!!!!!!!!!!!!!!!!!!!!!!
\subsection{Application of population synthesis}\label{SubSec:PSS}
% !!!!!!!!!!!!!!!!!!!!!!!!!!!!!!!!!!!!!!!!!!!!!!!!!!!!!!!!!!!!!!!!!!!!!!!!!!!!!!!!!!!!!!!!!!!!!!!!!!!!!!!!!!!!!!!!!!!!!!!!!!!!!!!!!!!!!!!!!!!!!!!!!!!!!!!!!!!!!!!!!!!!!!!!!!!!!!!!!!!!!!!!!!!!!!!

        Both purely stellar and active galaxy SEDs were modelled with the latest public distribution\footnote{{\sc Starlight}v04: \url{http://astro.ufsc.br/starlight/}} of \Starlight\ (\citealt{CidFernandes_etal_2005}) to investigate the ability of PSS to infer the SFHs and CEHs of mock galaxies. The code \Starlight\ is widely used (e.g.  \citealt{Mateus_etal_2006,  Stasinska_etal_2008, CidFernandes_etal_2010, CidFernandes_etal_2011, Kehrig_etal_2012, Benitez_etal_2013, Papaderos_etal_2013, CidFernandes_etal_2014, Stasinska_etal_2015, Gomes_etal_2016c}) and can  be regarded as a good representative of the state of the art. It is important to bear in mind the main objective of this work is to quantify potential biases on fundamental physical properties of active galaxies estimated with purely-stellar PSS that might rise owing to the introduction of a simple AGN model, as in \cite{Ciesla_etal_2015}.
        
\begin{figure*}[t]
\begin{center}
\includegraphics[width=1\textwidth]{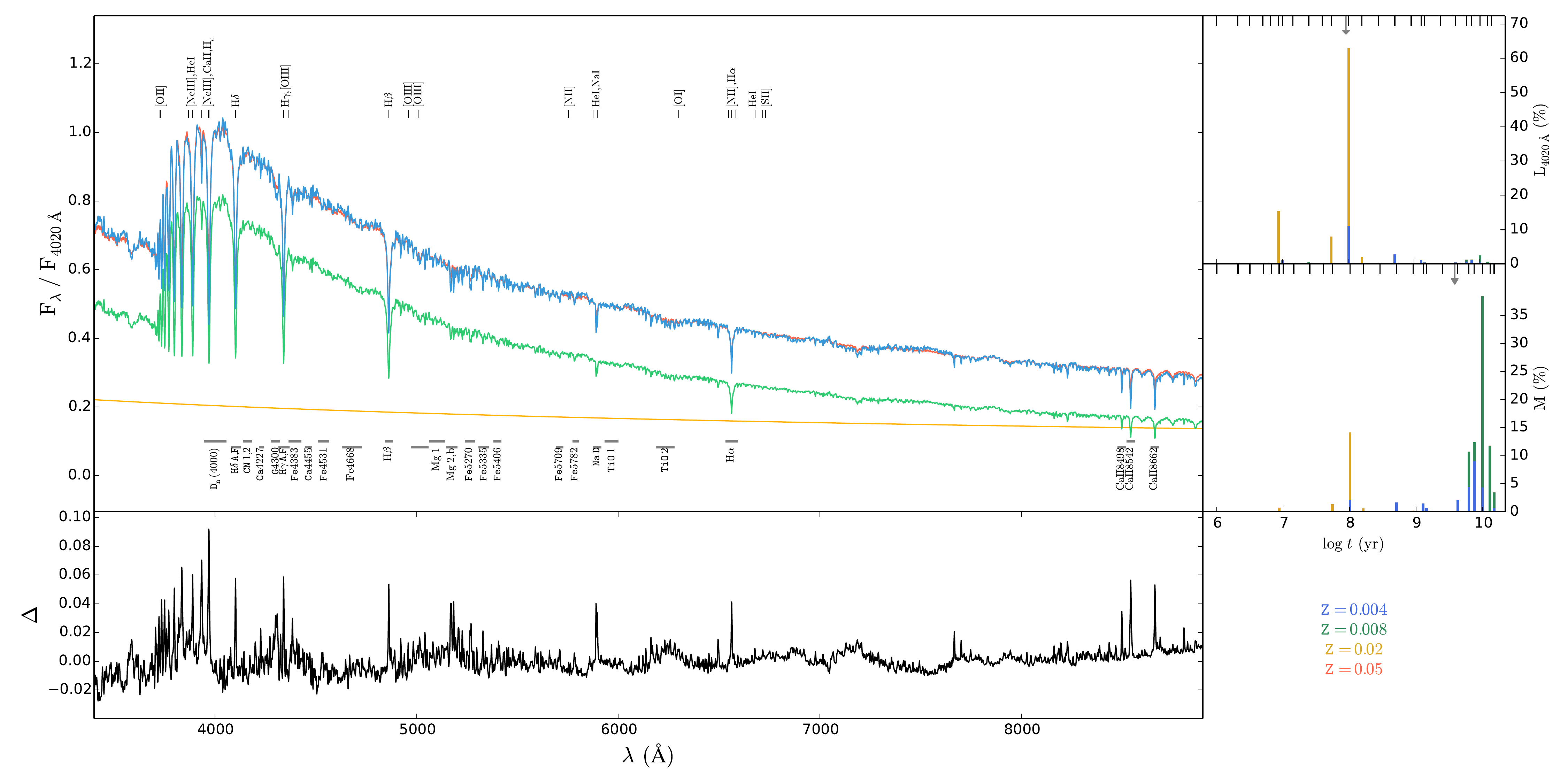}
\caption{Mean results of 10 \Starlight\ fits to a SED with a CSP with 100 Myr, solar metallicity, and an instantaneous burst SFH (green line) and an AGN PL with $\alpha=1.5$ and $x_{\mathrm{AGN}}=0.2$ (yellow line). Main panel: The red and cyan lines correspond to the input and and fitted spectra, respectively. The top and bottom annotations represent the wavelength range of several emission and absorption lines, respectively. Bottom panel: The residuals spectrum (black line) after subtracting the best fit from the input SED is shown. Top right panel: The SFHs in stellar light fractions is shown. Different colours correspond to different metallicities (see text below x-axis for label details). Lines on the top x-axis denote the age of the adopted base of SSPs. Bottom right panel: The SFHs in mass fractions are shown.  }
\label{Fig:Starligh_Fit_Example}
\end{center}
\end{figure*}

        The adopted base library comprises \cite{Bruzual_Charlot_2003} SSPs with 25 ages (illustrated  as the vertical black lines on the top of the right-hand side panels of Fig. \ref{Fig:Starligh_Fit_Example}) and 4 metallicities ($Z = 0.2$, 0.4, 1 and 2.5 $Z_{\odot}$). This library is similar to that adopted by \cite{Asari_etal_2007} with the following notable differences: first, better coverage of the younger evolutionary stages and, second, without the two lowest metallicities that are not expected to contribute significantly to the intrinsic degeneracies. A base library with a finer age and metallicity coverage would in principle help reduce uncertainties associated with poorly sampled evolutionary stages, although at a high computational cost. For instance, base libraries with 100 and 300 SSPs (i.e. the maximum number of base elements that \Starlight\ can handle) take $\sim$110 and 1000 seconds to perform a single fit to one of these synthetic spectra in a Intel\textsuperscript{\textregistered} Core\texttrademark ~~i7 CPU 870 @ 2.93GHz workstation desktop. Thus, the adopted base library should be regarded as a compromise between computational time and a fine age and metallicity coverage. The spectral fitting is made between 3400 and 8900 \AA, following common practice in local galaxy studies (e.g. \citealt{Asari_etal_2007, Ribeiro_etal_2016}) and to take advantage of the wavelength coverage of the STELIB stellar library (\citealt{LeBorgne_etal_2003}) adopted in the \cite{Bruzual_Charlot_2003} evolutionary models.

         Moreover, the stellar kinematics and V-band extinction $A_V$ were fixed to zero and no pixel clipping method was adopted. The reason for keeping stellar kinematics and extinction fixed to input values in the \Starlight\ fits is to facilitate an easier comparison between fits to purely stellar and active galaxy SEDs and to better isolate the potential PSS biases owing to the addition of the AGN PL continuum.

        The AGN models were processed by \Starlight\ for three fitting set-ups:
         
\begin{enumerate}
\item[(i)] without any PL contribution in the base
\item[(ii)] including in the base a single PL with same $\alpha$ as that embedded in the input spectra
\item[(iii)] including in the base PLs with $\alpha=0.5$, 1, 1.5 and 2, leaving to \Starlight\ the choice to pick up the PL which best matches the data.
\end{enumerate}
        
        Set-ups (\textit{i}) and (\textit{ii}) may be regarded as special cases in PSS modelling. The former corresponds to the worst-case scenario where fitting of an active galaxy SED is attempted with purely stellar templates, whereas the latter represents the ideal situation of the base library containing a single PL that is identical to that integrated in the input spectrum. Conversely, set-up (\textit{iii}) can be viewed as a conservative fitting approach in the situation of not having prior knowledge of the AGN optical continuum shape (e.g. \citealt{Ho_Kim_2009, Roche_etal_2016}).  Hereafter, these approaches are referred to as no power law (\npl), single power law (\spl), and multiple power laws (\mpl). In the following, focus is given to results derived for an instantaneous burst SFH. However, similar trends are seen for the other SFHs, results for which are discussed when appropriate.

%- - - - - - - - - - - - - - - - - - - - - - - - - - - - - - - - - - - - - - - - - - - - - - - - - - - - - - - - - - - - - - - - - - - - - - - - - - - - - - - - - - - - - - - - - - - - - 
%- - - - - - - - - - - - - - - - - - - - - - - - - - - - - - - - - - - - - - - - - - - - - - - - - - - - - - - - - - - - - - - - - - - - - - - - - - - - - - - - - - - - - - - - - - - - - 

        As an example, Figure \ref{Fig:Starligh_Fit_Example} shows the average results for 10 \Starlight\ fits for \npl\ with the best fit (blue line) to the input synthetic active galaxy spectrum (red line) with an stellar population instantaneously formed with 100 Myr and solar metallicity (green line) and an AGN PL with $\alpha=1.5$ and $x_{\mathrm{AGN}}=0.2$ (yellow line) in the main panel. The bottom panel presents the residuals after subtracting the fit from the input spectrum and the right-hand side panels show the estimated SFHs in terms of the light and mass fraction (top and bottom, respectively).  
                
% !!!!!!!!!!!!!!!!!!!!!!!!!!!!!!!!!!!!!!!!!!!!!!!!!!!!!!!!!!!!!!!!!!!!!!!!!!!!!!!!!!!!!!!!!!!!!!!!!!!!!!!!!!!!!!!!!!!!!!!!!!!!!!!!!!!!!!!!!!!!!!!!!!!!!!!!!!!!!!!!!!!!!!!!!!!!!!!!!!!!!!!!!!!!!!!
% - - - - - - - - - - - - - - - - -- - - - - - - -  - -  - -  - - RESULTS   - - - - - - - - - - - - - - - - - - - - - - - - - - - - - - - - - - - - - - - - - 
% !!!!!!!!!!!!!!!!!!!!!!!!!!!!!!!!!!!!!!!!!!!!!!!!!!!!!!!!!!!!!!!!!!!!!!!!!!!!!!!!!!!!!!!!!!!!!!!!!!!!!!!!!!!!!!!!!!!!!!!!!!!!!!!!!!!!!!!!!!!!!!!!!!!!!!!!!!!!!!!!!!!!!!!!!!!!!!!!!!!!!!!!!!!!!!!
\section{Results}\label{Sec:Results}

% Tau1 SFH = Mstellar vs model age
%- - - - - - - - - - - - - - - - - - - - - - - - - - - - - - - - - - - - - - - - - - - - - - - - - - - - - - - - - - - - - - - - - - - - - - - - - - - - - - - - - - - - - - - - - - - - - 
\begin{figure*}
\begin{center}
\includegraphics[width=0.95\textwidth]{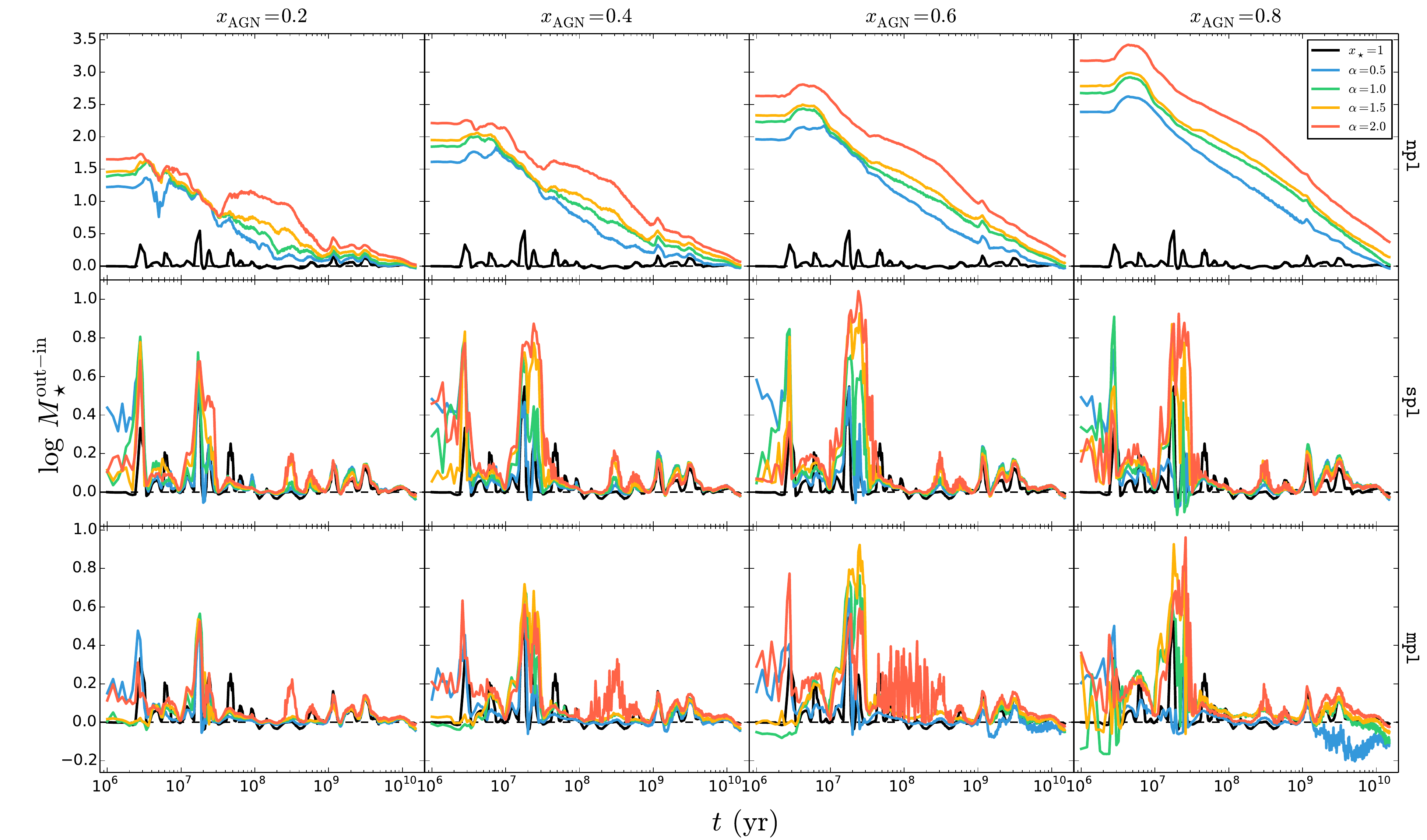}
\caption{Difference in the total stellar mass $M_{\star}$ between PSS (out) and ESS (in) values as a function of model age $t$ for a instantaneous burst SFH. Results adopting a base library without any AGN PL (\npl),  with a PL with the same index value as the input values (\spl) and with PLs with $\alpha=0.5$, 1, 1.5, and 2 (\mpl) are presented on top, middle, and bottom rows, respectively. Increasing AGN flux fractional contribution $x_{\mathrm{AGN}}$ at $\lambda_0=4020$ \AA\ are represented  from {left-} to {right-hand side panels}. Results from purely stellar models and models with PLs with $\alpha = 0.5$, 1.0, 1.5, and 2.0 are represented by {black}, {blue}, {green}, {yellow,} and {red lines}, respectively. }
\label{Fig:Tau1_-_Mavail_diff_vs_age}
\end{center}
\end{figure*}
%- - - - - - - - - - - - - - - - - - - - - - - - - - - - - - - - - - - - - - - - - - - - - - - - - - - - - - - - - - - - - - - - - - - - - - - - - - - - - - - - - - - - - - - - - - - - - 

        Figure \ref{Fig:Tau1_-_Mavail_diff_vs_age} shows the ratio in the currently available total stellar mass $M_{\star}$ between the PSS (output) and ESS (input) values as a function of age $t$, with increasing $x_{\mathrm{AGN}}$ from left- to right-hand side panels, respectively. The results from \npl, \spl,\ and \mpl\ are presented on the top, middle, and bottom rows, respectively (subsequent figures follow a similar panel configuration, except when explicitly stated otherwise). Moreover, PSS results for purely stellar spectra results are represented by black lines, while active galaxy spectra with $\alpha=0.5$, 1, 1.5, and 2 are represented by blue, green, yellow, and red lines, respectively.  The results for purely stellar models show a maximum mass overestimation up to $\sim$0.5 dex, which are likely linked to a poor age coverage of the SSPs in the adopted base library.
                
        The results in Fig. \ref{Fig:Tau1_-_Mavail_diff_vs_age} for \npl\ show a systematic mass overestimation up to $\sim$3.5 dex correlated with increasing $\alpha$ and $x_{\mathrm{AGN}}$ and with decreasing age $t$, with a maximum at $t\sim10$ Myr. Figure \ref{Fig:All_SFHs_-_Mstar_diff_vs_age} in Appendix \ref{Appendix:Additional_Resources} shows similar results for other SFHs.  Results for \spl\ and \mpl\ reveal local mass overestimations within $\sim$1 dex. This overestimation increases with increasing $x_{\mathrm{AGN}}$, is somewhat independent from $\alpha$ and becomes more severe in models with ages where spectral synthesis of purely stellar spectra also show considerable uncertainties.

        One question of considerable interest concerns the origin of this severe stellar mass overestimation. Results show that \Starlight\ has to compensate for the lack of the PL component in the base in a purely stellar modelling configuration. This leads to a nearly bimodal mixture of young and old stellar populations, as illustrated in the SFH panels of Fig. \ref{Fig:Starligh_Fit_Example}.  It is to be expected that in such a modelling approach the blue AGN continuum is accounted for by a non-realistic mixture of stellar populations, where very young SSPs would always dominate the light, whereas the contribution of old SSPs is almost negligible in terms of light. Notwithstanding this fact, these old SSPs provide the bulk of stellar mass, thus leading to the observed mass overestimation. 

% Tau1 SFH = logt_L vs model age
%- - - - - - - - - - - - - - - - - - - - - - - - - - - - - - - - - - - - - - - - - - - - - - - - - - - - - - - - - - - - - - - - - - - - - - - - - - - - - - - - - - - - - - - - - - - - - 
\begin{figure*}
\begin{center}
\includegraphics[width=0.95\textwidth]{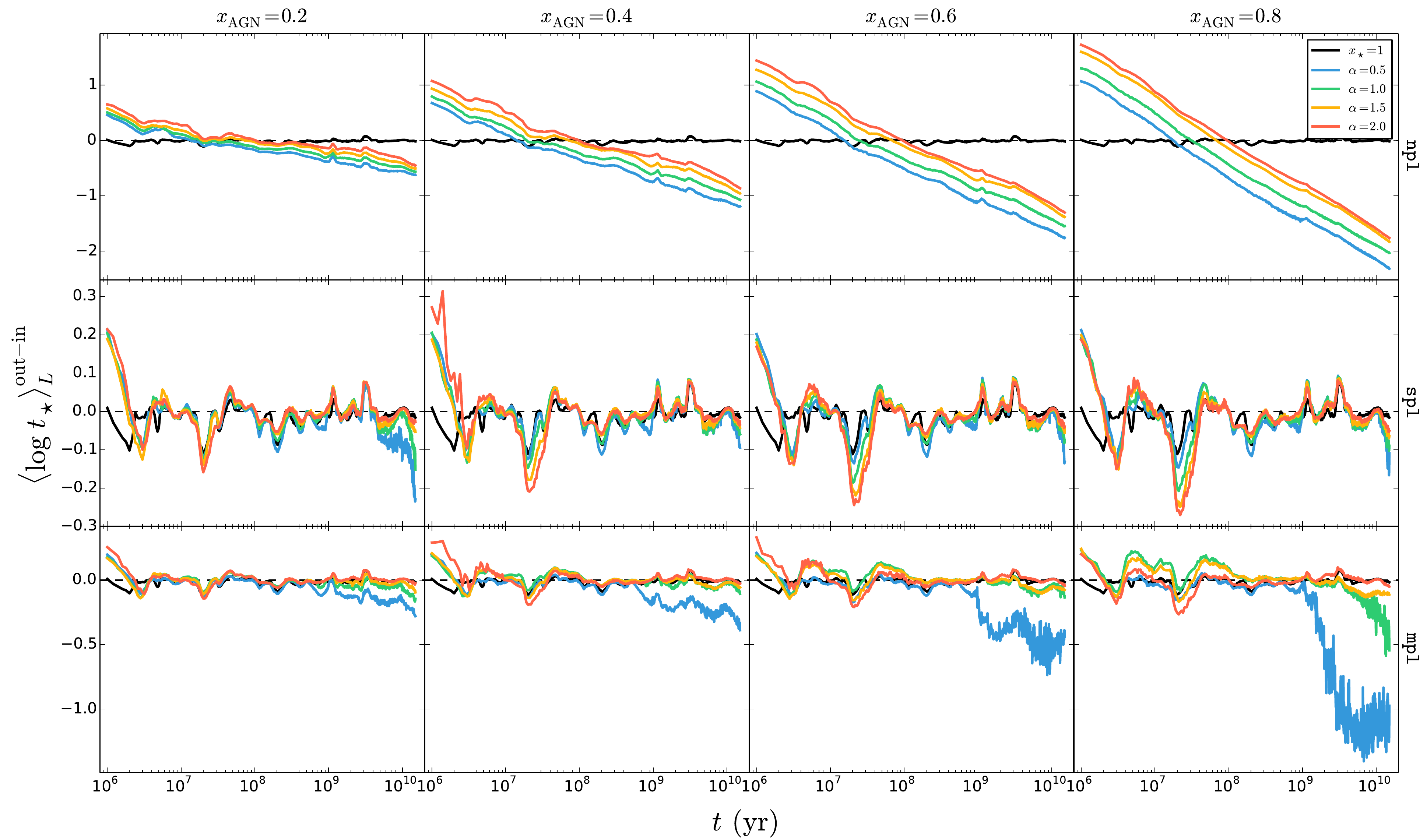}
\caption{Difference in the light-weighted mean stellar age $ \langle \log t_{\star} \rangle_L$ between PSS and ESS values as a function of age $t$ for an instantaneous burst SFH. Panel configuration and legend details are analogous to those of Fig. \ref{Fig:Tau1_-_Mavail_diff_vs_age}.}
\label{Fig:Tau1_-_logtL_diff_vs_age}
\end{center}
\end{figure*}
%- - - - - - - - - - - - - - - - - - - - - - - - - - - - - - - - - - - - - - - - - - - - - - - - - - - - - - - - - - - - - - - - - - - - - - - - - - - - - - - - - - - - - - - - - - - - - 

        Figure \ref{Fig:Tau1_-_logtL_diff_vs_age} shows the difference in the light-weighted mean stellar age $\langle \log t_{\star} \rangle_L$ between PSS and ESS values as a function of age $t$. The results of purely stellar models show mean stellar age uncertainties within $\sim$0.1 dex. \cite{CidFernandes_etal_2005} showed that both light-weighted mean stellar age and metallicity could be recovered with \Starlight\ within a typical uncertainty of $\sim$0.2 dex using mock data with S/N=10. This uncertainty is considered in this work to be a fiducial value.

        The results in Fig. \ref{Fig:Tau1_-_logtL_diff_vs_age} for \npl\ show a systematic age over- and underestimation of up to $\sim$1.5 and $\sim$2 dex for young ($t \lesssim 100$ Myr) and old ($t \gtrsim 100$ Myr) evolutionary models, respectively. This trend is accentuated with increasing $x_{\mathrm{AGN}}$  and originates from an increasing dilution of stellar absorption features, as seen in Fig. \ref{Fig:Dilution}.  Figure \ref{Fig:All_SFHs_-_logt_L_difference_vs_age} shows similar results for other SFHs. Moreover,  the results for \spl\ show an age over- and underestimation of up to $\sim$0.3 dex, which appear to be correlated with ages where purely stellar fits also show considerable uncertainties. In addition, there is a clear age underestimation at $\sim$20 Myr that increases with increasing $\alpha$ and $x_{\mathrm{AGN}}$. The results for \mpl\ show uncertainties and trends similar to those found for \spl. However, the age of old evolutionary models ($t>1$ Gyr) can be underestimated by as much as $\sim$1.5 dex. This feature becomes more prominent with increasing $x_{\mathrm{AGN}}$ for $\alpha=0.5$. The results in Fig \ref{Fig:Tau1_-_x_AGN_diff_vs_age} suggest that this trend is due to a poor estimation of $x_{\mathrm{AGN}}$ for this fitting set-up, which also explains the slight mass underestimation for this age and AGN contribution.
        
% Tau1 SFH = logt_M vs model age
%- - - - - - - - - - - - - - - - - - - - - - - - - - - - - - - - - - - - - - - - - - - - - - - - - - - - - - - - - - - - - - - - - - - - - - - - - - - - - - - - - - - - - - - - - - - - - 
\begin{figure*}
\begin{center}
\includegraphics[width=0.95\textwidth]{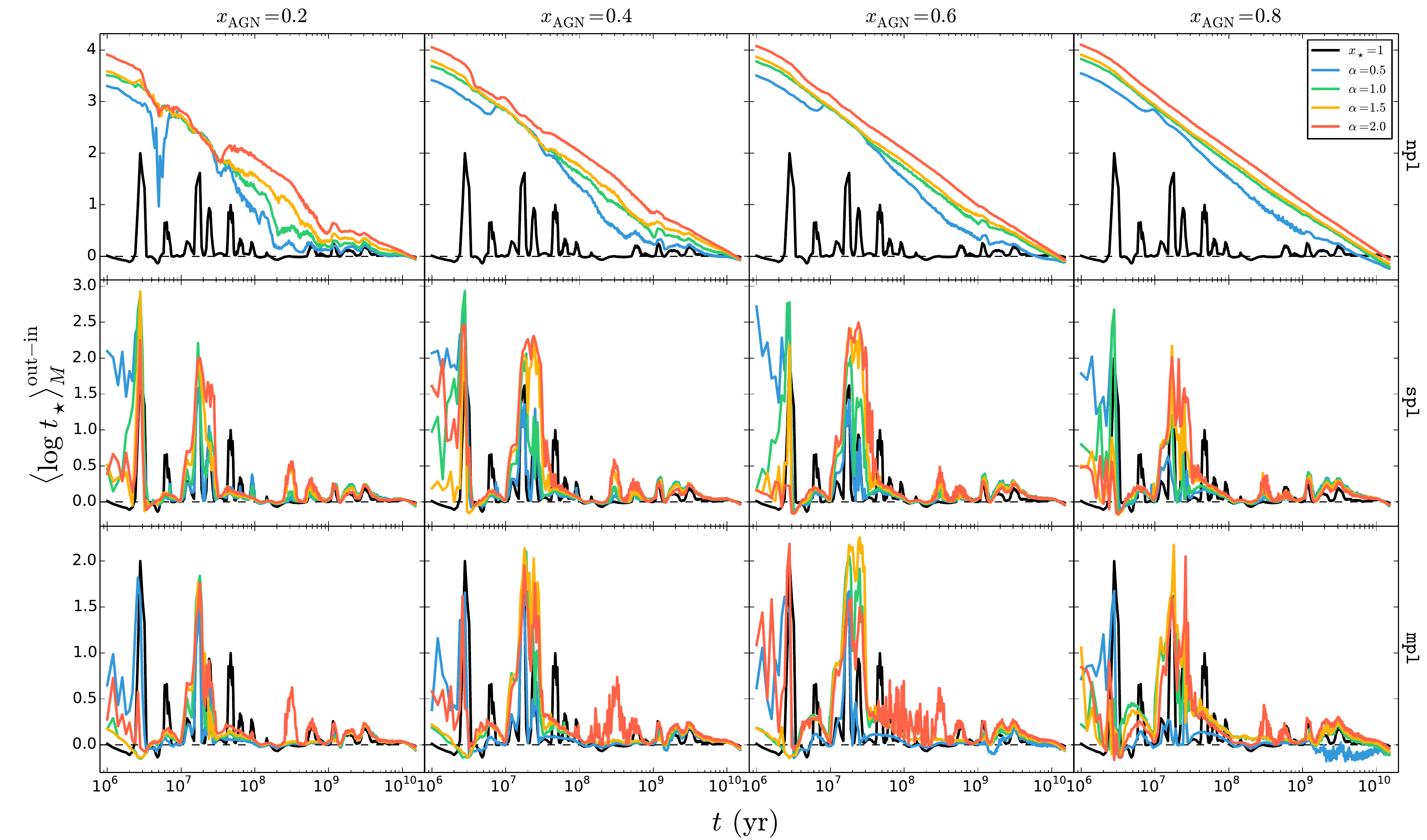}
\caption{Difference in the mass-weighted mean stellar age $ \langle \log t_{\star} \rangle_M$ between PSS and ESS values as a function of age $t$ for an instantaneous burst SFH. Panel configuration and legend details are analogous to those of Fig. \ref{Fig:Tau1_-_Mavail_diff_vs_age}.}
\label{Fig:Tau1_-_logtM_diff_vs_age}
\end{center}
\end{figure*}
%- - - - - - - - - - - - - - - - - - - - - - - - - - - - - - - - - - - - - - - - - - - - - - - - - - - - - - - - - - - - - - - - - - - - - - - - - - - - - - - - - - - - - - - - - - - - - 

        Figure \ref{Fig:Tau1_-_logtM_diff_vs_age} illutrating results for mass-weighted mean stellar age $\langle \log t_{\star} \rangle_M$  show overall uncertainties that are larger than their light-weighted counterparts. Synthesis results of purely stellar models show a maximum age overestimation up to $\sim$2 dex. Moreover, the results for \npl\ show a systematic age overestimation up to $\sim$4 dex with $\alpha$ and decreasing $t$. Figure \ref{Fig:All_SFHs_-_logt_M_difference_vs_age} shows similar results for other SFHs. This trend is similar to that found in the top row of Fig. \ref{Fig:Tau1_-_Mavail_diff_vs_age}. Results for \spl\ and \mpl\ show age overestimations up to $\sim$3 and 2 dex, respectively,  in ages where purely stellar models display considerable uncertainties, similarly to total stellar mass results in Fig. \ref{Fig:Tau1_-_Mavail_diff_vs_age}. 

% Tau1 SFH = logZ_L vs model age        
%- - - - - - - - - - - - - - - - - - - - - - - - - - - - - - - - - - - - - - - - - - - - - - - - - - - - - - - - - - - - - - - - - - - - - - - - - - - - - - - - - - - - - - - - - - - - - 
\begin{figure*}
\begin{center}
\includegraphics[width=0.95\textwidth]{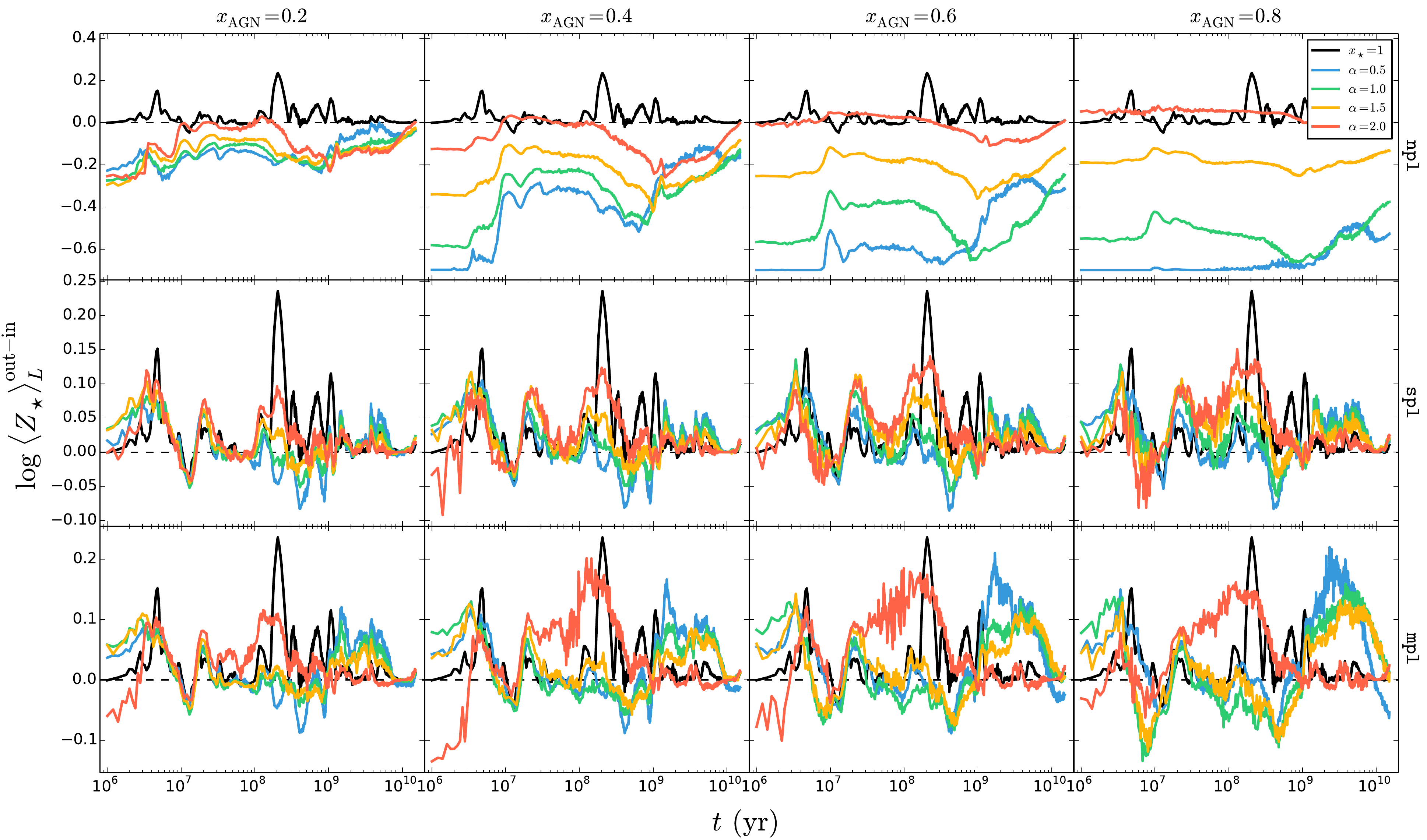}
\caption{Difference in the light-weighted mean stellar metallicity $\log \langle Z_{\star} \rangle_L$ between PSS and ESS values as a function of age $t$ for an instantaneous burst SFH. Panel configuration and legend details are analogous to those of Fig. \ref{Fig:Tau1_-_Mavail_diff_vs_age}.}
\label{Fig:Tau1_-_logZL_diff_vs_age}
\end{center}
\end{figure*}
%- - - - - - - - - - - - - - - - - - - - - - - - - - - - - - - - - - - - - - - - - - - - - - - - - - - - - - - - - - - - - - - - - - - - - - - - - - - - - - - - - - - - - - - - - - - - -
        
        Figure \ref{Fig:Tau1_-_logZL_diff_vs_age} shows the difference in the light-weighted mean stellar metallicity $\log  \langle Z_{\star} \rangle_L$ between PSS and ESS values as a function of age $t$. The results for the purely stellar models show metallicity over- and underestimations within $\sim$0.2 dex correlated with evolutionary ages with light-weighted age under- and overestimations, respectively. This is likely due to age-metallicity degeneracies that prevent a clear distinction between young metal-rich and old metal-poor stellar populations or vice versa (\citealt{Faber_1972, OConnell_1980, Renzini_Buzzoni_1986, Bressan_Chiosi_Tantalo_1996, Pelat_1997, Pelat_1998, CidFernandes_etal_2005}). This problem arises after changes of absorption line features induced by variations in both age and metallicity (e.g. \citealt{Worthey_1994}), which is aggravated in this case by the dilution of the absorption features by the PL. 
        
        The results in Fig. \ref{Fig:Tau1_-_logZL_diff_vs_age} for \npl\ show a systematic underestimation of metallicity by up to $\sim$0.7 dex with decreasing $\alpha$ and age $t$ and increasing $x_{\mathrm{AGN}}$. Figure \ref{Fig:All_SFHs_-_logZ_L_difference_vs_age} shows similar results for other SFHs.  The metallicity underestimation plateau at -0.7 dex seen for $\alpha=2$ and $x_{\mathrm{AGN}}=0.8$ when $t<1$ Gyr corresponds to a metallicity of $0.2 \, Z_{\odot}$, which is the lowest metallicity found in the base. Thus, this result is due to the construction of the base library and has no physical meaning. In addition, this result is due to an increasing dilution of the absorption features with increasing $x_{\mathrm{AGN}}$, which artificially makes the stellar continuum increasingly metal poorer. The results for \spl\ display random metallicity uncertainties up to $\sim$0.15 dex predominately located around ages where purely stellar synthesis results also show considerable uncertainties. These roughly increase with increasing $\alpha$ and $x_{\mathrm{AGN}}$. Moreover, the results for \mpl\ show an increasing metallicity overestimation up to $\sim$0.2 dex with increasing $\alpha$ and $x_{\mathrm{AGN}}$ for $t\gtrsim1$ Gyr, mirroring the age underestimation at the same evolutionary stages shown in Fig. \ref{Fig:Tau1_-_logtL_diff_vs_age}.

% Tau1 SFH = logZ_M vs model age        
%- - - - - - - - - - - - - - - - - - - - - - - - - - - - - - - - - - - - - - - - - - - - - - - - - - - - - - - - - - - - - - - - - - - - - - - - - - - - - - - - - - - - - - - - - - - - - 
\begin{figure*}
\begin{center}
\includegraphics[width=0.95\textwidth]{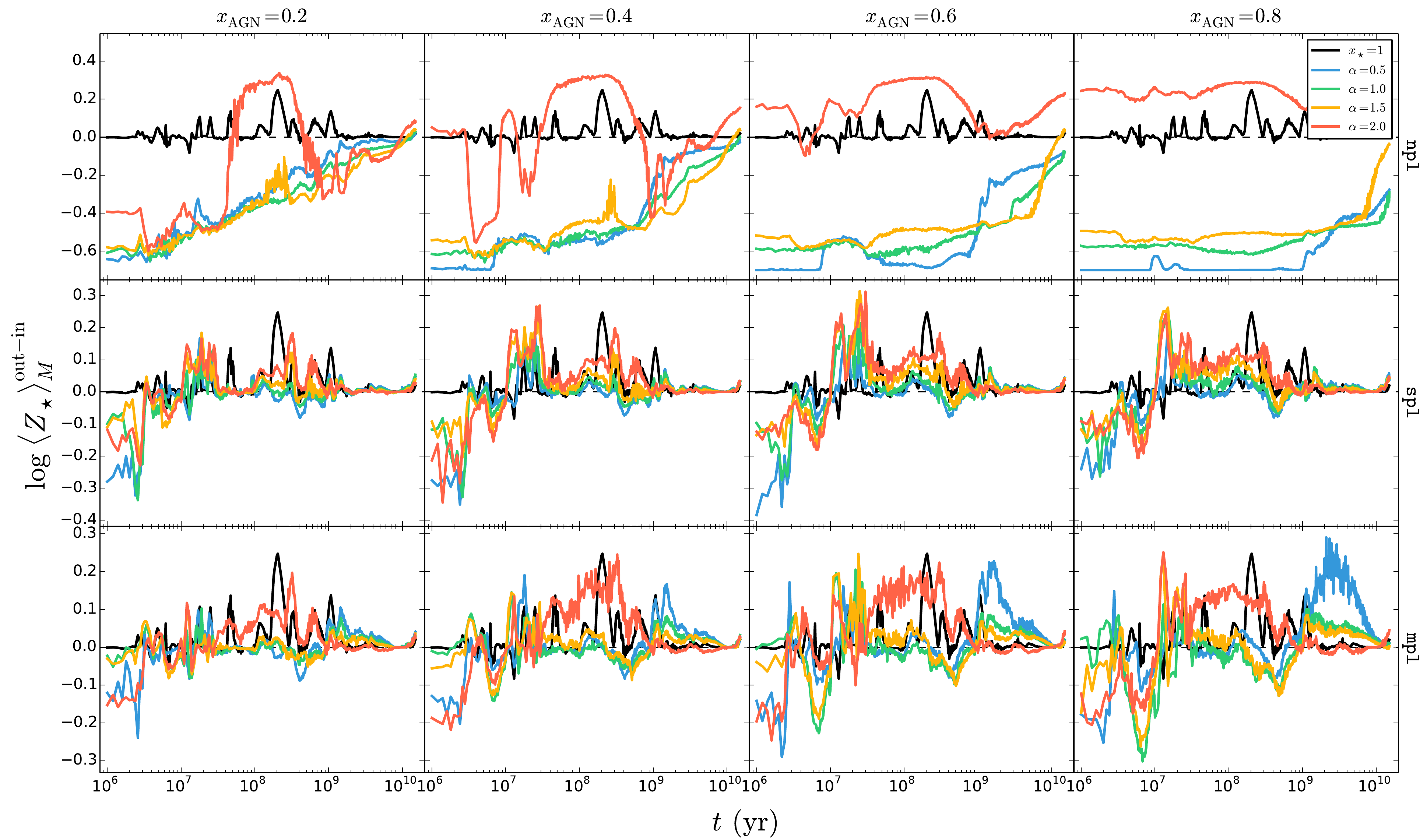}
\caption{Difference in the mass-weighted mean stellar metallicity $\log \langle Z_{\star} \rangle_M$ between PSS and ESS values as a function of age $t$ for an instantaneous burst SFH. Panel configuration and legend details are analogous to those of Fig. \ref{Fig:Tau1_-_Mavail_diff_vs_age}.}
\label{Fig:Tau1_-_logZM_diff_vs_age}
\end{center}
\end{figure*}
%- - - - - - - - - - - - - - - - - - - - - - - - - - - - - - - - - - - - - - - - - - - - - - - - - - - - - - - - - - - - - - - - - - - - - - - - - - - - - - - - - - - - - - - - - - - - - 

        Figure \ref{Fig:Tau1_-_logZM_diff_vs_age} illustrating results for the mass-weighted mean stellar metallicity $\log  \langle Z_{\star} \rangle_M$ show that light- and mass-weighted metallicities display similar trends with $t$, $\alpha$ and $x_{\mathrm{AGN}}$ for all set-ups. Indeed, the results for \npl\ show again that the metallicity can be underestimated by up to $\sim$0.7 dex. Figure \ref{Fig:All_SFHs_-_logZ_L_difference_vs_age} shows similar results for other SFHs. Results also show that the mass-weighted properties tend to have larger uncertainties than their light-weighted counterparts. The reason for this is that small light variations translate into large mass variations when applying the mass-to-light ratio to the light fractions vector of the SSPs to determine mass related properties.

        Figures \ref{Fig:Cont_-_Mavail_diff_vs_age}--\ref{Fig:Cont_-_logZM_diff_vs_age} in Appendix \ref{Appendix:Additional_Resources} show plots of spectral synthesis results for a continuous SFH  analogous to those presented in Figures \ref{Fig:Tau1_-_Mavail_diff_vs_age}--\ref{Fig:Tau1_-_logZM_diff_vs_age}. These results show that stellar uncertainties induced by the added AGN PL component are relatively lower for a continuous SFH than those estimated for an instantaneous burst SFH. This happens because short phases of significant spectrophotometric evolution in the first few hundred Myr are largely washed out in the case of continuous star formation when looking at the time evolution of light- and mass-weighted stellar properties. With respect to an instantaneous burst of a given age and metallicity, the SSP that best approximates this evolutionary stage is not necessarily included in the base library. In this case, and assuming the best case scenario in which there are almost no degeneracies among the stellar populations, adjacent SSPs in age and/or metallicity would have to be used, which unavoidably induces strong local deviations between the true and estimated evolutionary quantities.  Thus, results obtained for the instantaneous burst SFH should be viewed as upper limits to the uncertainties on the analysed stellar properties. 
        
% !!!!!!!!!!!!!!!!!!!!!!!!!!!!!!!!!!!!!!!!!!!!!!!!!!!!!!!!!!!!!!!!!!!!!!!!!!!!!!!!!!!!!!!!!!!!!!!!!!!!!!!!!!!!!!!!!!!!!!!!!!!!!!!!!!!!!!!!!!!!!!!!!!!!!!!!!!!!!!!!!!!!!!!!!!!!!!!!!!!!!!!!!!!!!!!
% - - - - - - - - - - - - - - - - -- - - - - - - -  - -  - - - - - DISCUSSION   - - - - - - - - - - - - - - - - - - - - - - - - - - - - - - - - - - - - - -
% !!!!!!!!!!!!!!!!!!!!!!!!!!!!!!!!!!!!!!!!!!!!!!!!!!!!!!!!!!!!!!!!!!!!!!!!!!!!!!!!!!!!!!!!!!!!!!!!!!!!!!!!!!!!!!!!!!!!!!!!!!!!!!!!!!!!!!!!!!!!!!!!!!!!!!!!!!!!!!!!!!!!!!!!!!!!!!!!!!!!!!!!!!!!!!!
\section{Discussion}\label{Sec:Discussion}

        A substantial body of work devoted to the exploration of the assembly history of galaxies relies on automated PSS modelling of large extragalactic data sets (e.g. \citealt{Kauffmann_etal_2003c,Asari_etal_2007,CidFernandes_etal_2010}).  A common practice in these studies is the prior exclusion from the PSS modelling of galaxies classified as Seyfert on the basis of BPT emission-line diagnostics, since both the spectroscopic characteristics of these systems (e.g. broad emission lines and dilution of stellar features by the featureless AGN PL) and code-specific limitations generally hinder a reliable separation of the non-thermal and stellar SEDs and the extraction from the latter of key physical and evolutionary properties (e.g. stellar mass, mean stellar age and metallicity,  and SFH).  However, as pointed out in \citet{Papaderos_etal_2013} and \citet{Gomes_etal_2016c} (see also CGP16), standard BPT classification diagnostics become inapplicable in the case of virtually gas-evacuated galaxies where the bulk of the Lyman continuum radiation from an AGN eventually escapes without locally producing detectable optical line emission. This, together with dilution of nuclear emission-line EWs by the stellar component along the line of sight, may readily prevent detection of accretion-powered nuclear activity in an early-type galaxy (\citealt{Papaderos_etal_2013,Gomes_etal_2016c}), which is then  classified as {retired/passive} and included in automated PSS studies.  The same may obviously apply to old classical bulges, many of which show faint nebular emission despite hosting a super-massive black hole  whose accretion-powered energy release may result in significant contamination of the optical spectrum by an AGN PL component. Our pilot study in CGP16 has first addressed the question of the detectability of an AGN PL embedded within an old instantaneously formed Lyman-photon leaking galaxy\ and explored the effect that this PL component may have on optical PSS modelling studies. This analysis has revealed that a PL contributing up to $\sim$26\% ($\equiv x_{\rm AGN}^{\rm threshold}$) of the monochromatic luminosity at 4020 \AA\ generally evades detection both from visual inspection and PSS modelling of a galaxy spectrum and introduces a substantial bias in the physical properties of the stellar component obtained through spectral synthesis. Here, we  extend this study by modelling with the PSS code \Starlight\ an extensive grid of synthetic stellar SEDs that trace the spectral evolution of galaxies for a wide range of SFHs over an age span between 1 Myr and 15 Gyr. A comparison of the physical properties obtained for the stellar component from PSS modelling with the input values constrained by the synthetic SEDs has allowed to examine in detail potential biases in PSS modelling for different fitting set-ups (Sect.~\ref{Sec:Results}).

        Of special importance is the fact that, regardless of the spectral index $\alpha$ of the PL component integrated in the synthetic input SEDs, PSS fitting with purely stellar templates yields at $x_{\rm AGN}^{\rm threshold}$ an overestimation of stellar mass $M_{\star}$ by up to two orders of magnitude for instantaneously formed stellar populations of age $\la 10^8$ yr. Even though this bias is becoming gradually smaller with increasing age of the input SEDs, it still impacts $M_{\star}$ estimates by a factor of $\sim$2 over several Gyr of galactic evolution. For the same set of synthetic SEDs, PSS fits overestimate (underestimate) the light-weighted stellar age by up to $\sim$1 dex of  stellar populations younger (older) than 10 Myr, whereas the mass-weighted stellar age is overestimated by up to $\sim$3 dex for young ages and a factor $\geq$2 for old ($\sim$5 Gyr) stellar populations.  As for the mass-weighted stellar metallicity, our study indicates a systematic underestimation by a factor of $\sim$3 for young ages with this bias decreasing yet still present beyond an age of $\sim$1 Gyr. Such biases are also well documented for SFHs involving a prolonged star formation process (e.g. continuous or exponentially decreasing), suggesting a principal trend for purely stellar PSS fits to compensate for the presence of a PL spectral component through a severely overestimated contribution from old, low-metallicity stars.

        Regarding the impact of an AGN PL on the stellar continuum, stellar population biases correlate on first order with $x_{\mathrm{AGN}}$ and on second with $\alpha$. These are mainly due to the combined effect of the dilution of stellar absorption features by the AGN continuum and a degeneracy between the AGN PL and young SSPs.  The addition of a PL leads to a stellar age underestimation given the decrease of the D$_n$4000 \AA\ break strength and simultaneously to a stellar metallicity overestimation due to the dilution of Balmer absorption features. Moreover, Balmer absorption features become weaker with age and their EWs are inversely correlated with D$_n$4000 for continuous SFHs (\citealt{Kauffmann_etal_2003c}).

        A question that naturally arises in view of these modelling biases pertains to the interpretation of the nature of evolved Lyman-photon-leaking galaxies hosting an AGN. Taken at face value, PSS fits might prompt the conclusion that these systems are extraordinarily massive and have assembled quasi-monolithically early on, which would be consistent with the much lower level of chemical enrichment as compared to local galaxies of equal $M_{\star}$. A possibly significant population of such old, ultra-massive galaxies erroneously included in galaxy statistics could then impact determinations of the cosmic SFH.
 
% !!!!!!!!!!!!!!!!!!!!!!!!!!!!!!!!!!!!!!!!!!!!!!!!!!!!!!!!!!!!!!!!!!!!!!!!!!!!!!!!!!!!!!!!!!!!!!!!!!!!!!!!!!!!!!!!!!!!!!!!!!!!!!!!!!!!!!!!!!!!!!!!!!!!!!!!!!!!!!!!!!!!!!!!!!!!!!!!!!!!!!!!!!!!!!!
% - - - - - - - - - - - - - - - - -- - - - - - - -  - -  - - CONCLUSIONS   - - - - - - - - - - - - - - - - - - - - - - - - - - - - - - - - - - - - - - -
% !!!!!!!!!!!!!!!!!!!!!!!!!!!!!!!!!!!!!!!!!!!!!!!!!!!!!!!!!!!!!!!!!!!!!!!!!!!!!!!!!!!!!!!!!!!!!!!!!!!!!!!!!!!!!!!!!!!!!!!!!!!!!!!!!!!!!!!!!!!!!!!!!!!!!!!!!!!!!!!!!!!!!!!!!!!!!!!!!!!!!!!!!!!!!!!
\section{Conclusions}\label{Sec:Conclusions}

        Synthetic spectra of active galaxies for solar metallicity and multiple SFHs were created with the ESS code \Rebetiko\ and used to investigate how a simple AGN FC model impacts the estimation of stellar population properties with the state-of-the-art PSS code \Starlight. The AGN continuum model is defined by its PL index $\alpha$ and AGN fractional contribution $x_{\mathrm{AGN}}$ to the $\lambda_0=4020$ \AA\ monochromatic flux.  The stellar spectral contribution is computed assuming solar metallicity SSPs from \cite{Bruzual_Charlot_2003} in both the creation of the synthetic spectra with ESS and the application of PSS. The accuracy of the estimated physical properties depends on the model age, $\alpha$, $x_{\mathrm{AGN}}$ and, more importantly, the approach adopted for dealing with the AGN continuum:
        
\begin{enumerate}
\item Excluding any PL component in the fit can lead to uncertainties on stellar mass by up to $\sim$3.5 dex, light- and mass-weighted mean stellar age up to $\sim$2 and $\sim$4 dex, respectively, and on both light- and mass-weighted mean stellar metallicity up to $\sim$0.7 dex.
        
\item Including a single or multiple PL components in the fit leads to uncertainties on stellar mass overestimation up to $\sim$1 dex, light- and -mass-weighted mean stellar age up to $\sim$1.5 and $\sim$3 dex, respectively, and light- and mass-weighted mean stellar metallicity up to $\sim$0.2 and $\sim$0.4 dex, respectively.
\end{enumerate}

        The uncertainties on the stellar population properties estimated with PSS are weakly dependent on the SFH, in particular when no AGN model is included in the base of elements. These results might lead to the misinterpretation that evolved Lyman-photon-leaking galaxies hosting an AGN are particularly massive, metal poor, and formed monolithically when the universe was still very young. Hence, these results show the importance of accounting for AGN spectral contributions when applying state-of-the-art PSS to active galaxies for a viable estimation of their physical properties, such as star formation and chemical evolution histories.

% !!!!!!!!!!!!!!!!!!!!!!!!!!!!!!!!!!!!!!!!!!!!!!!!!!!!!!!!!!!!!!!!!!!!!!!!!!!!!!!!!!!!!!!!!!!!!!!!!!!!!!!!!!!!!!!!!!!!!!!!!!!!!!!!!!!!!!!!!!!!!!!!!!!!!!!!!!!!!!!!!!!!!!!!!!!!!!!!!!!!!!!!!!!!!!!
% - - - - - - - - - - - - - - - - -- - - - - - - - - - - - - ACKNOWLEDGEMENTS   - - - - - - - - - - - - - - - - - - - - - - - - - - - - - - - 
% !!!!!!!!!!!!!!!!!!!!!!!!!!!!!!!!!!!!!!!!!!!!!!!!!!!!!!!!!!!!!!!!!!!!!!!!!!!!!!!!!!!!!!!!!!!!!!!!!!!!!!!!!!!!!!!!!!!!!!!!!!!!!!!!!!!!!!!!!!!!!!!!!!!!!!!!!!!!!!!!!!!!!!!!!!!!!!!!!!!!!!!!!!!!!!!

\begin{acknowledgements}
    
	This work was supported by Funda\c{c}\~{a}o para a Ci\^{e}ncia e a Tecnologia (FCT) through national funds (UID/FIS/04434/2013) and by FEDER through COMPETE2020 (POCI-01-0145-FEDER-007672). We acknowledge support from the FCT through national funds (PTDC/FIS-AST/3214/2012) and by FEDER through COMPETE (FCOMP-01-0124-FEDER-029170). 
	We thank the anonymous referee for valuable comments and suggestions.
	LSMC acknowledges funding by FCT through the grant CIAAUP-01/2016-BI in the context of the FCT project UID/FIS/04434/2013 \& POCI-01-0145-FEDER-007672. 
	JMG is supported by the fellowship SFRH/BPD/66958/2009 funded by FCT (Portugal) and POPH/FSE (EC) and by the fellowship CIAAUP-04/2016-BPD in the context of the FCT project UID/FIS/04434/2013 \& POCI-01-0145-FEDER-007672. 
	PP is supported by FCT through Investigador FCT contract IF/01220/2013/CP1191/CT0002 and by a Ci\^encia 2008 contract, funded by FCT/MCTES (Portugal) and POPH/FSE (EC).	
	We also acknowledge the exchange programme ``Study of Emission-Line Galaxies with Integral-Field Spectroscopy'' (SELGIFS, FP7-PEOPLE-2013-IRSES-612701), funded by the EU through the IRSES scheme.

\end{acknowledgements}

% !!!!!!!!!!!!!!!!!!!!!!!!!!!!!!!!!!!!!!!!!!!!!!!!!!!!!!!!!!!!!!!!!!!!!!!!!!!!!!!!!!!!!!!!!!!!!!!!!!!!!!!!!!!!!!!!!!!!!!!!!!!!!!!!!!!!!!!!!!!!!!!!!!!!!!!!!!!!!!!!!!!!!!!!!!!!!!!!!!!!!!!!!!!!!!!
% - - - - - - - - - - - - - - - - -- - - - - - - - - - - - - - - - - BIBLIOGRAPHY   - - - - - - - - - - - - - - - - - - - - - - - - - - - - - - -  - - -
% !!!!!!!!!!!!!!!!!!!!!!!!!!!!!!!!!!!!!!!!!!!!!!!!!!!!!!!!!!!!!!!!!!!!!!!!!!!!!!!!!!!!!!!!!!!!!!!!!!!!!!!!!!!!!!!!!!!!!!!!!!!!!!!!!!!!!!!!!!!!!!!!!!!!!!!!!!!!!!!!!!!!!!!!!!!!!!!!!!!!!!!!!!!!!!!
%\include{References}

\normalsize
% !!!!!!!!!!!!!!!!!!!!!!!!!!!!!!!!!!!!!!!!!!!!!!!!!!!!!!!!!!!!!!!!!!!!!!!!!!!!!!!!!!!!!!!!!!!!!!!!!!!!!!!!!!!!!!!!!!!!!!!!!!!!!!!!!!!!!!!!!!!!!!!!!!!!!!!!!!!!!!!!!!!!!!!!!!!!!!!!!!!!!!!!!!!!!!!
% - - - - - - - - - - - - - - - - -- - - - - - - - - - - - - - - - -  - - - APPENDIX   - - - - - - - - - - - - - - - - - - - - - - - - - - - - - - -  - - - - 
% !!!!!!!!!!!!!!!!!!!!!!!!!!!!!!!!!!!!!!!!!!!!!!!!!!!!!!!!!!!!!!!!!!!!!!!!!!!!!!!!!!!!!!!!!!!!!!!!!!!!!!!!!!!!!!!!!!!!!!!!!!!!!!!!!!!!!!!!!!!!!!!!!!!!!!!!!!!!!!!!!!!!!!!!!!!!!!!!!!!!!!!!!!!!!!!
\appendix
\section[]{Additional resources}\label{Appendix:Additional_Resources}

%% Tau1 SFH = Delta_x_AGN vs model age
%- - - - - - - - - - - - - - - - - - - - - - - - - - - - - - - - - - - - - - - - - - - - - - - - - - - - - - - - - - - - - - - - - - - - - - - - - - - - - - - - - - - - - - - - - - - - - 
\begin{figure*}
\begin{center}
\includegraphics[width=0.95\textwidth]{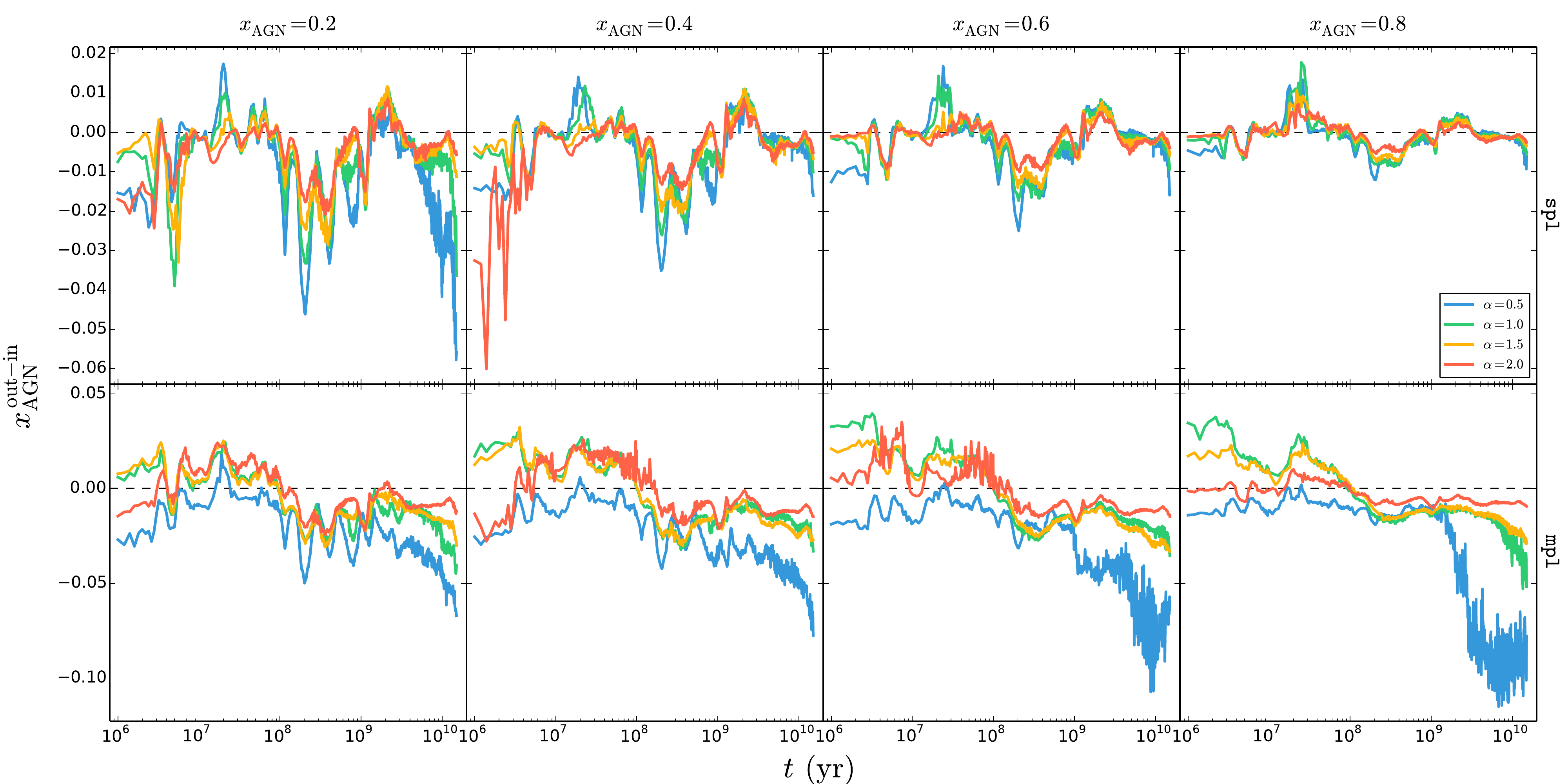}
\caption{Difference in AGN fractional contribution $x_{\mathrm{AGN}}$ between PSS and ESS values as a function of model age $t$ for an instantaneous burst SFH with increasing AGN fractional contribution $x_{\mathrm{AGN}}$ from left- to right-hand side panels. Top and bottom rows illustrate results for \spl\ and \mpl, respectively. }
\label{Fig:Tau1_-_x_AGN_diff_vs_age}
\end{center}
\end{figure*}
%- - - - - - - - - - - - - - - - - - - - - - - - - - - - - - - - - - - - - - - - - - - - - - - - - - - - - - - - - - - - - - - - - - - - - - - - - - - - - - - - - - - - - - - - - - - - - 

\begin{figure*}
\begin{center}
\includegraphics[width=0.95\textwidth]{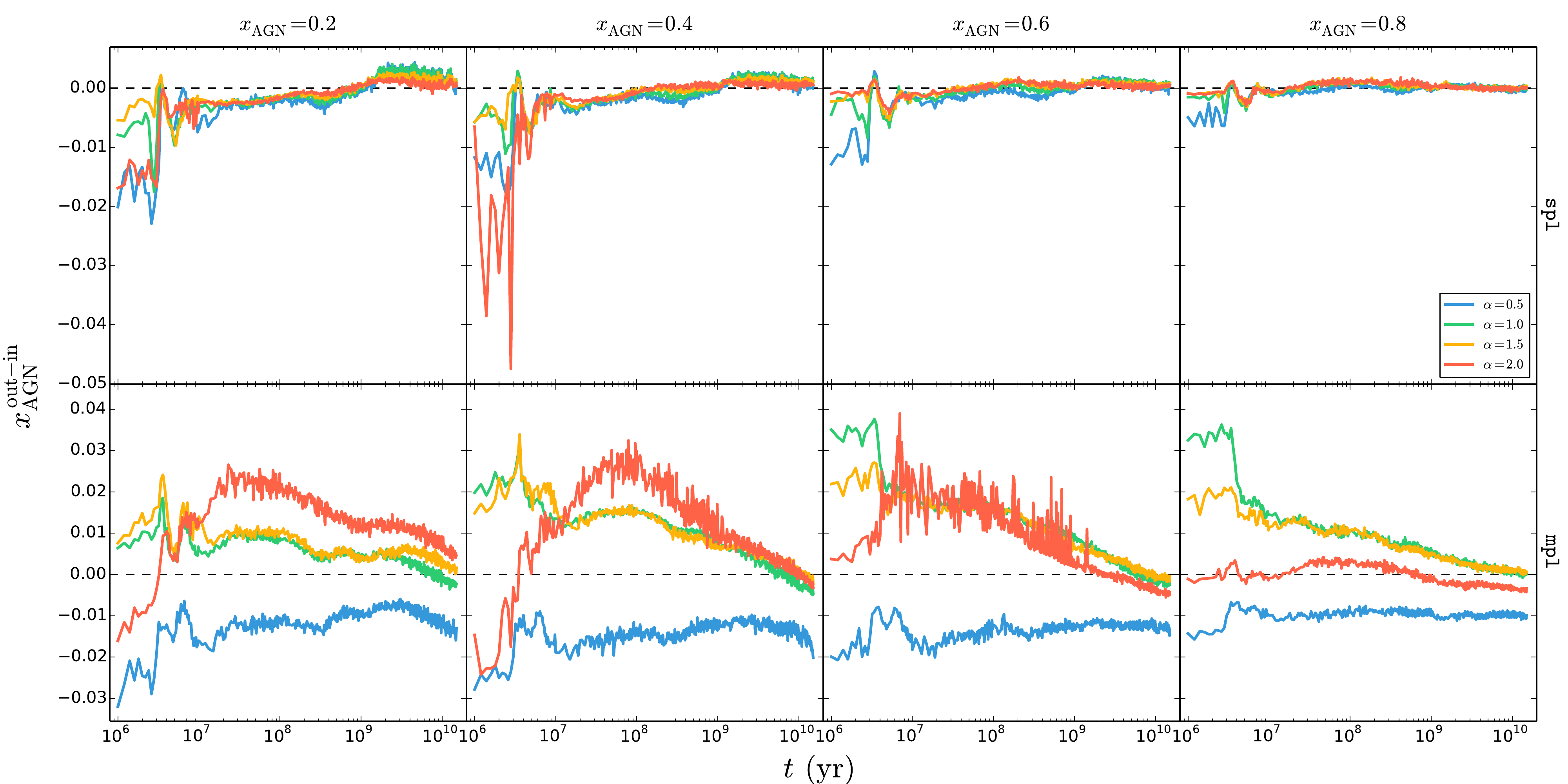}
\caption{Difference in the AGN fractional contribution $x_{\mathrm{AGN}}$ between PSS and ESS values as a function of model age $t$ for a continuous SFH. Panel configuration and legend details are analogous to those of Fig. \ref{Fig:Tau1_-_x_AGN_diff_vs_age}.}
\label{Fig:Cont_-_x_AGN_diff_vs_age}
\end{center}
\end{figure*}

        Figures \ref{Fig:Tau1_-_x_AGN_diff_vs_age} and \ref{Fig:Cont_-_x_AGN_diff_vs_age} show the AGN fractional contribution $x_{\mathrm{AGN}}$ difference between PSS and ESS as a function of model age $t$ for models with instantaneous burst and continuous SFHs, respectively, with increasing $x_{\mathrm{AGN}}$ from left- to -right-hand side panels. Top and bottom row panels show results for the \spl\ and \mpl\ fitting set-ups (Section \ref{SubSec:PSS}). On the one hand, the results for \spl\ show that the accuracy of the estimated $x_{\mathrm{AGN}}$ increases with $\alpha$ and $x_{\mathrm{AGN}}$. Moreover, there is a underestimation bump between $10^8$ and $10^9$ yr that is attenuated with increasing $x_{\mathrm{AGN}}$. On the other hand, the results for \mpl\ show a systematic $x_{\mathrm{AGN}}$ under- and overestimation for $t\lesssim10^8$ and $t\gtrsim10^8$ yr with increasing $x_{\mathrm{AGN}}$, respectevely. This underestimation increases with decreasing $\alpha$. Results for both \spl\ and \mpl\ show that the estimation of $x_{\mathrm{AGN}}$ is sensitive to the underlying stellar continuum shape.

% All SFHs = Mstellar vs model age for all SFHs
%- - - - - - - - - - - - - - - - - - - - - - - - - - - - - - - - - - - - - - - - - - - - - - - - - - - - - - - - - - - - - - - - - - - - - - - - - - - - - - - - - - - - - - - - - - - - 
\begin{figure*}
\includegraphics[width=1\textwidth]{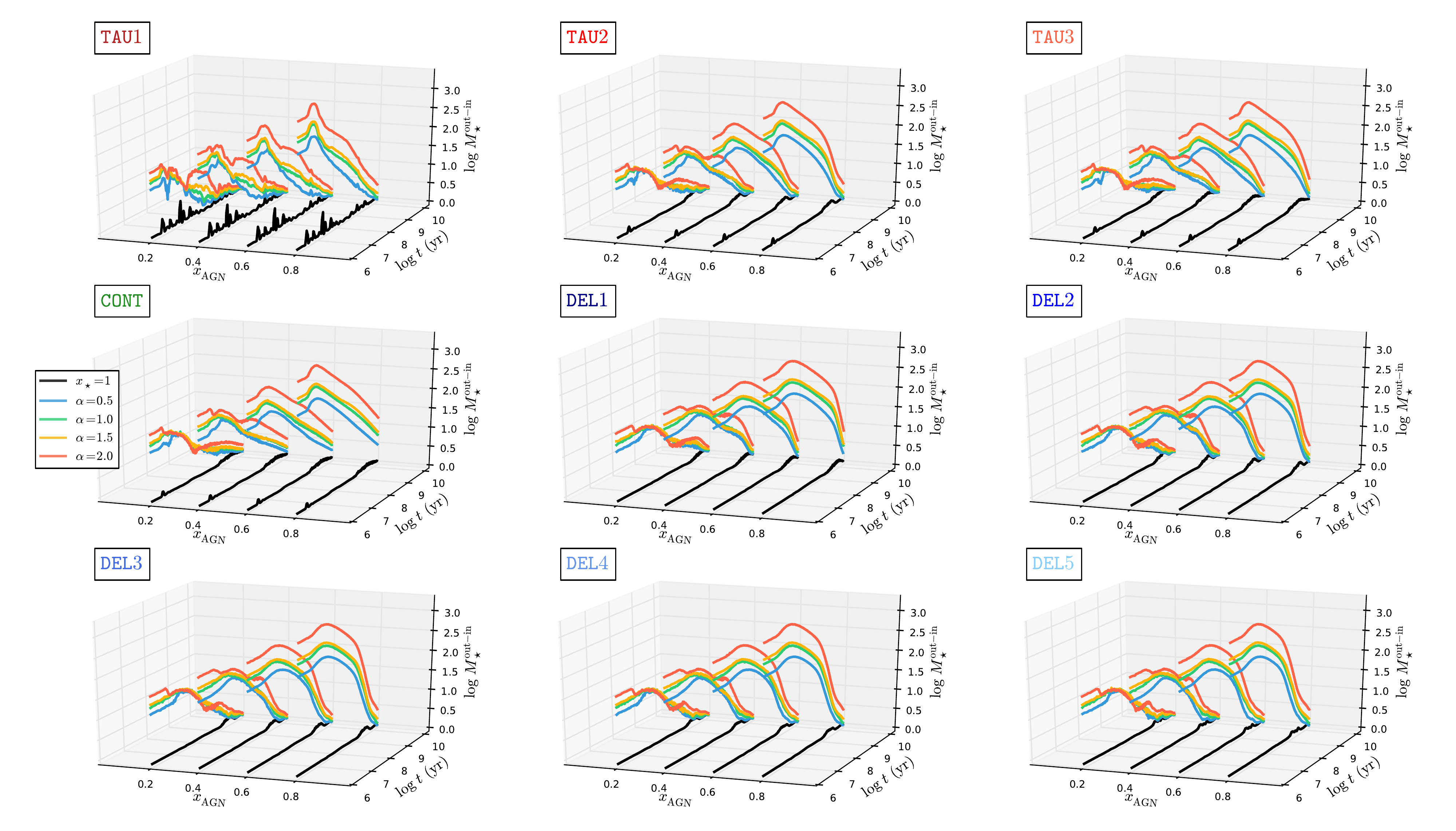}
\caption{Difference in the stellar mass  $M_{\star}$ (z-axis) between PSS and ESS values as a function of model age $t$ (y-axis) and AGN fractional contribution $x_{\mathrm{AGN}}$ (x-axis).  Black, blue, green, yellow, and red lines represent CSP models and active galaxy models with $\alpha = 0.5$, 1.0, 1.5, and 2.0, respectively. Each panel corresponds to different SFHs (see Fig. \ref{Fig:SFRs_of_REBETIKO} for label details).}
\label{Fig:All_SFHs_-_Mstar_diff_vs_age}
\end{figure*}
%- - - - - - - - - - - - - - - - - - - - - - - - - - - - - - - - - - - - - - - - - - - - - - - - - - - - - - - - - - - - - - - - - - - - - - - - - - - - - - - - - - - - - - - - - - - - 
% All SFHs = logt_L vs model age for all SFHs
%- - - - - - - - - - - - - - - - - - - - - - - - - - - - - - - - - - - - - - - - - - - - - - - - - - - - - - - - - - - - - - - - - - - - - - - - - - - - - - - - - - - - - - - - - - - - 
\begin{figure*}
\includegraphics[width=1\textwidth]{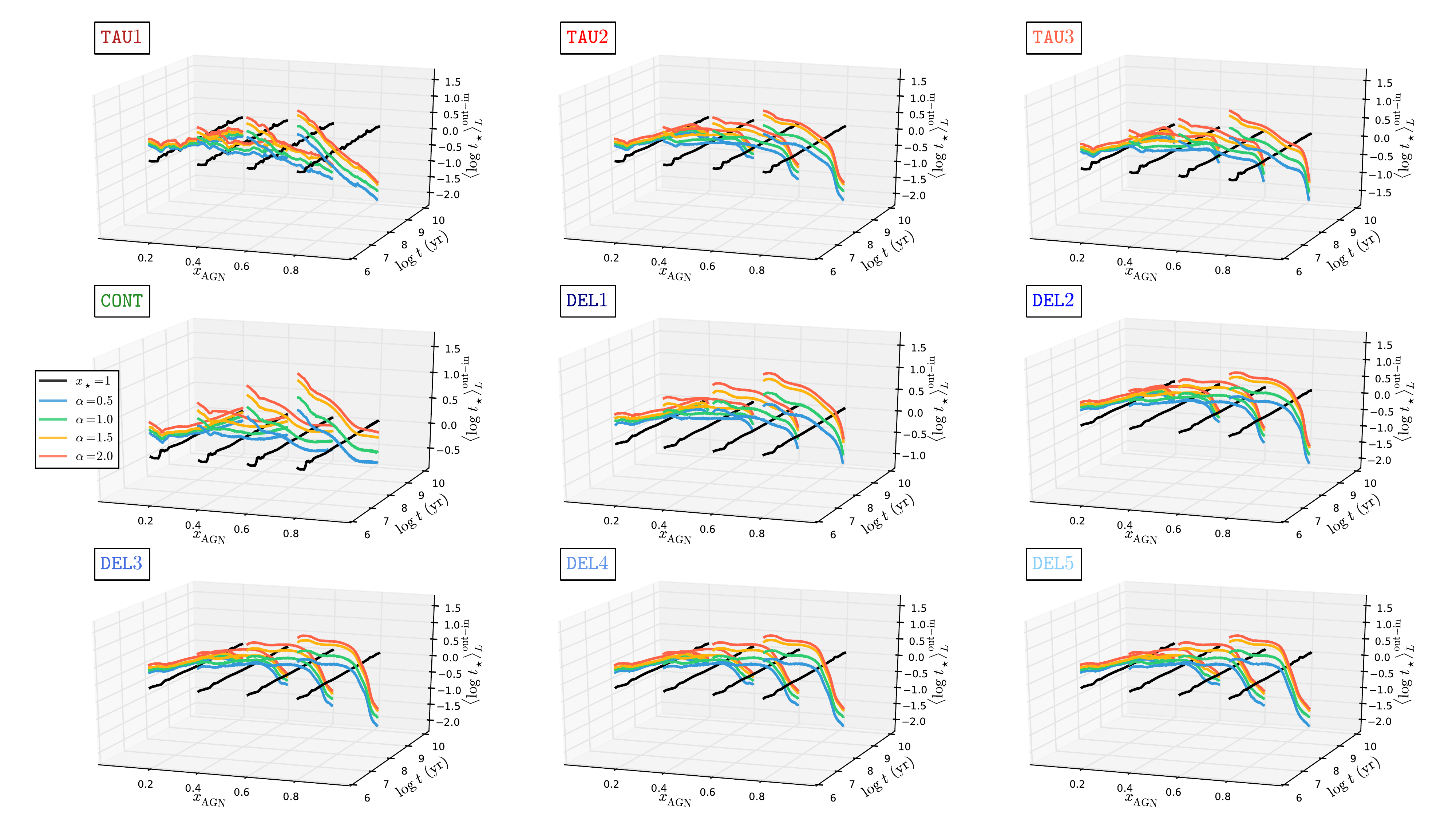}
\caption{Difference in the light-weighted mean stellar age $ \langle \log t_{\star} \rangle_L$ (z-axis) between PSS and ESS values as a function of model age $t$ (y-axis) and AGN fractional contribution (x-axis). Panel configuration and legend details are analogous to those of Fig. \ref{Fig:All_SFHs_-_Mstar_diff_vs_age}.}
\label{Fig:All_SFHs_-_logt_L_difference_vs_age}
\end{figure*}
%- - - - - - - - - - - - - - - - - - - - - - - - - - - - - - - - - - - - - - - - - - - - - - - - - - - - - - - - - - - - - - - - - - - - - - - - - - - - - - - - - - - - - - - - - - - - 
% All SFHs = logt_M vs model age for all SFHs
%- - - - - - - - - - - - - - - - - - - - - - - - - - - - - - - - - - - - - - - - - - - - - - - - - - - - - - - - - - - - - - - - - - - - - - - - - - - - - - - - - - - - - - - - - - - - 
\begin{figure*}
\includegraphics[width=1\textwidth]{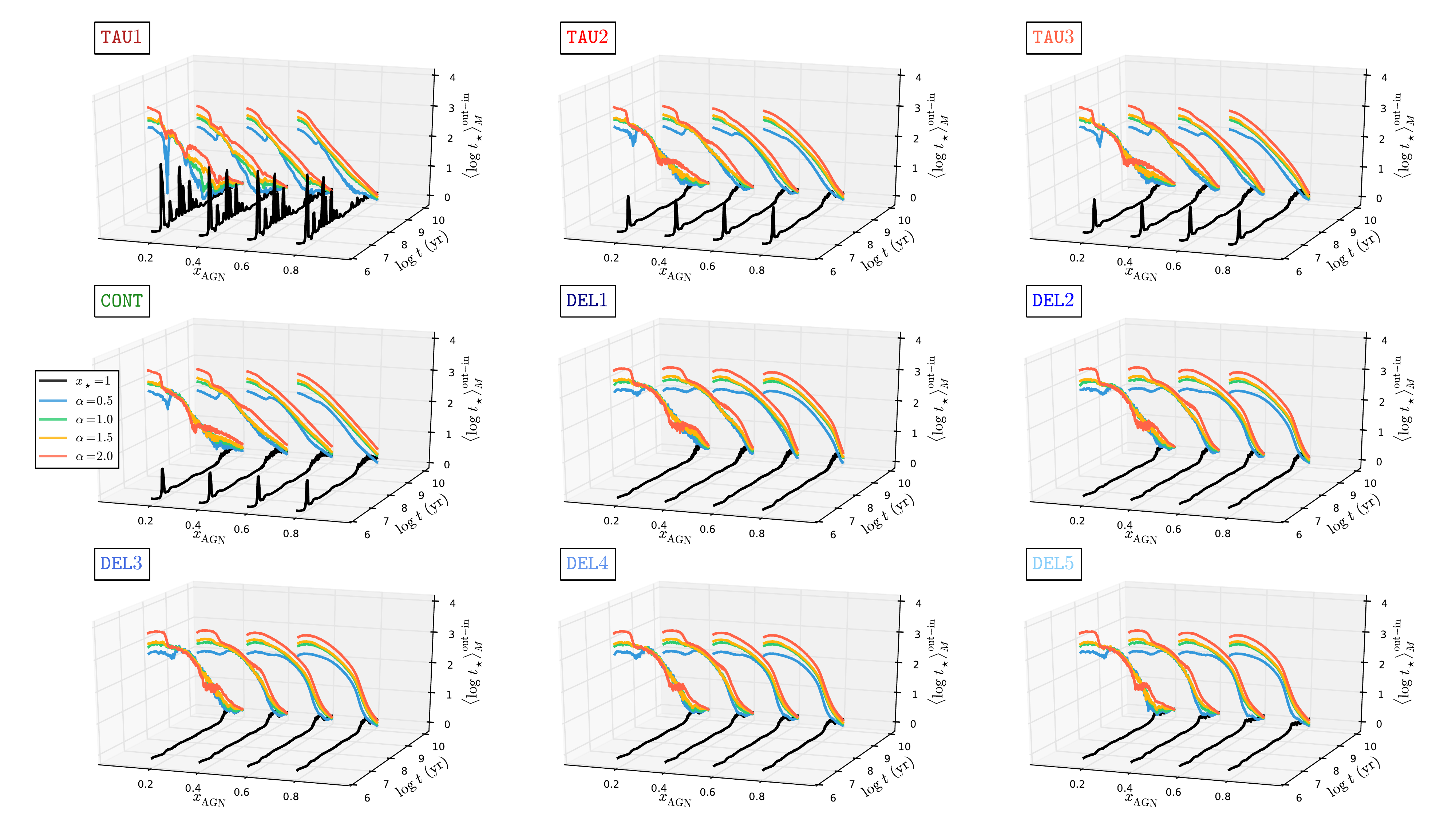}
\caption{Difference in the mass-weighted mean stellar age $ \langle \log t_{\star} \rangle_M$ (z-axis) between PSS and ESS values as a function of model age $t$ (y-axis) and AGN fractional contribution (x-axis). Panel configuration and legend details are analogous to those of Fig. \ref{Fig:All_SFHs_-_Mstar_diff_vs_age}.}
\label{Fig:All_SFHs_-_logt_M_difference_vs_age}
\end{figure*}
%- - - - - - - - - - - - - - - - - - - - - - - - - - - - - - - - - - - - - - - - - - - - - - - - - - - - - - - - - - - - - - - - - - - - - - - - - - - - - - - - - - - - - - - - - - - - 
% All SFHs = logZ_L vs model age for all SFHs
%- - - - - - - - - - - - - - - - - - - - - - - - - - - - - - - - - - - - - - - - - - - - - - - - - - - - - - - - - - - - - - - - - - - - - - - - - - - - - - - - - - - - - - - - - - - - 
\begin{figure*}
\includegraphics[width=1\textwidth]{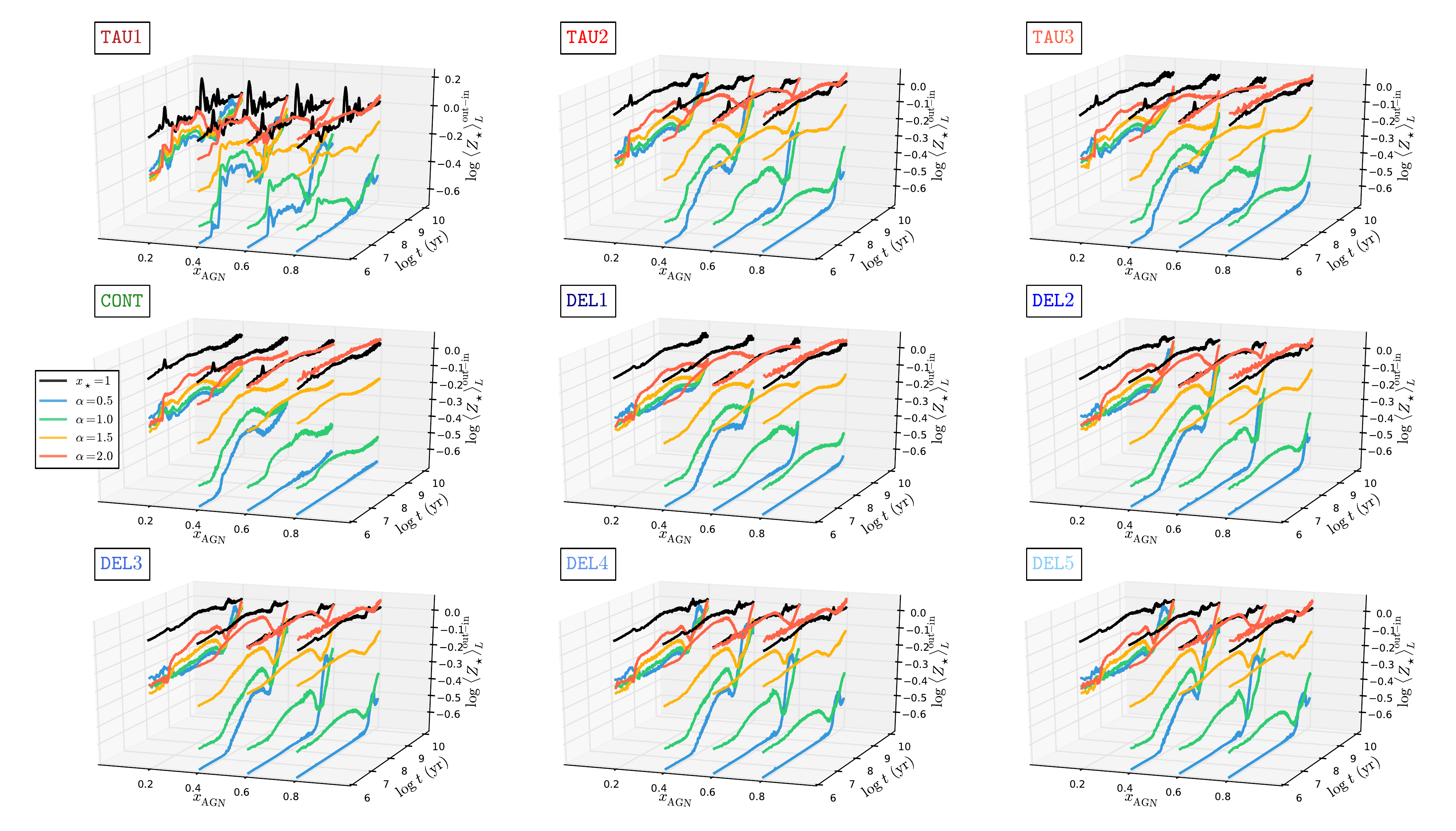}
\caption{Difference in the light-weighted mean stellar metallicity $\log \, \langle  Z_{\star} \rangle_L$ (z-axis) between PSS and ESS values as a function of model age $t$ (y-axis) and AGN fractional contribution (x-axis). Panel configuration and legend details are analogous to those of Fig. \ref{Fig:All_SFHs_-_Mstar_diff_vs_age}.}
\label{Fig:All_SFHs_-_logZ_L_difference_vs_age}
\end{figure*}
%- - - - - - - - - - - - - - - - - - - - - - - - - - - - - - - - - - - - - - - - - - - - - - - - - - - - - - - - - - - - - - - - - - - - - - - - - - - - - - - - - - - - - - - - - - - - 
% All SFHs = logZ_M vs model age for all SFHs
%- - - - - - - - - - - - - - - - - - - - - - - - - - - - - - - - - - - - - - - - - - - - - - - - - - - - - - - - - - - - - - - - - - - - - - - - - - - - - - - - - - - - - - - - - - - - 
\begin{figure*}
\includegraphics[width=1\textwidth]{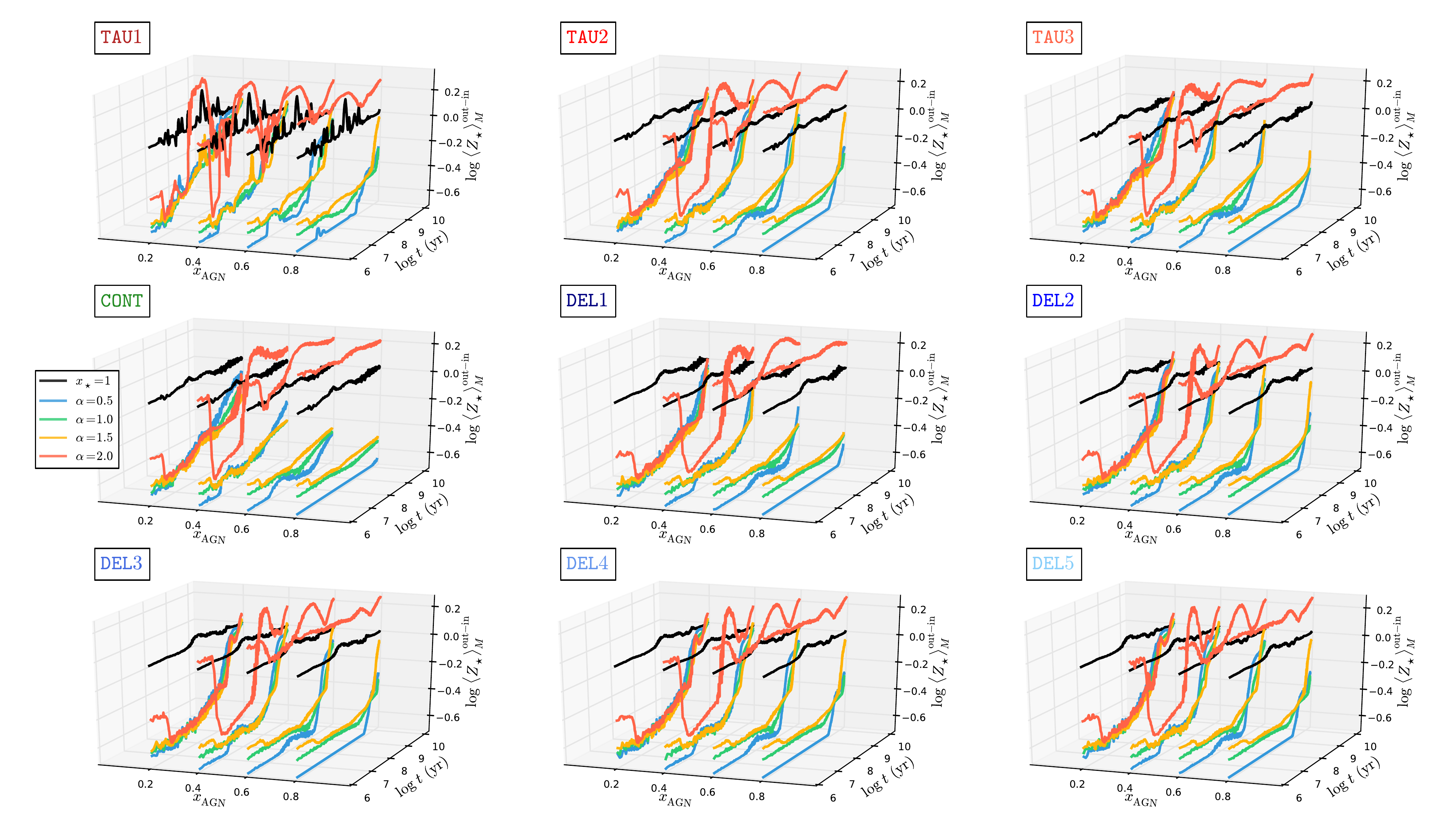}
\caption{Difference in the mass-weighted mean stellar metallicity $\log \, \langle  Z_{\star} \rangle_M$ (z-axis) between PSS and ESS values as a function of model age $t$ (y-axis) and AGN fractional contribution (x-axis). Panel configuration and legend details are analogous to those of Fig. \ref{Fig:All_SFHs_-_Mstar_diff_vs_age}.}
\label{Fig:All_SFHs_-_logZ_M_difference_vs_age}
\end{figure*}
%- - - - - - - - - - - - - - - - - - - - - - - - - - - - - - - - - - - - - - - - - - - - - - - - - - - - - - - - - - - - - - - - - - - - - - - - - - - - - - - - - - - - - - - - - - - - 

% Cont SFH = Mstellar vs model age
%- - - - - - - - - - - - - - - - - - - - - - - - - - - - - - - - - - - - - - - - - - - - - - - - - - - - - - - - - - - - - - - - - - - - - - - - - - - - - - - - - - - - - - - - - - - - 
\begin{figure*}
\begin{center}
\includegraphics[width=0.95\textwidth]{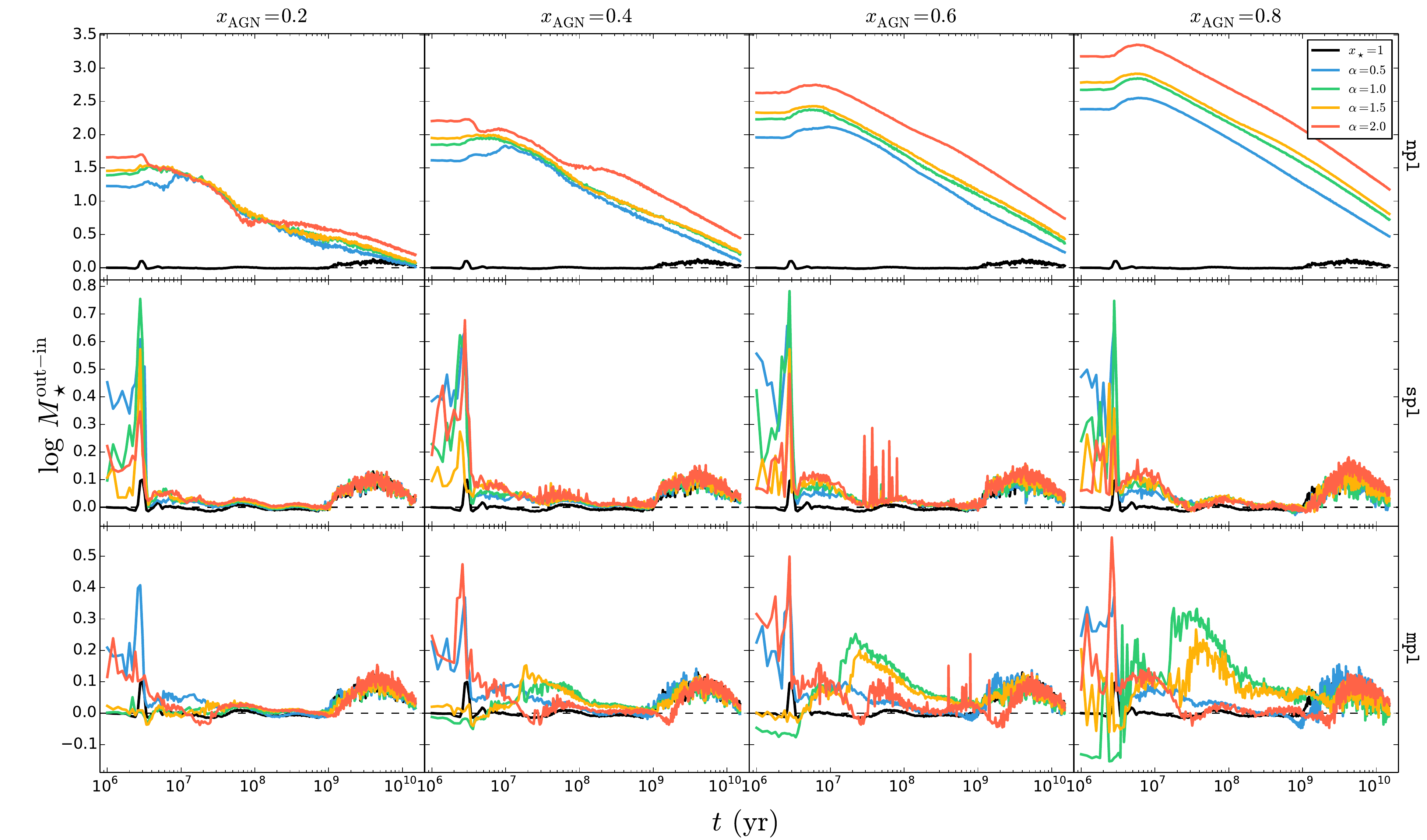}
\caption{Difference in the total stellar mass $M_{\star}$ between PSS and ESS values as a function of model age $t$ for a continuous SFH. Panel configuration and legend details are analogous to those of Fig. \ref{Fig:Tau1_-_Mavail_diff_vs_age}. }
\label{Fig:Cont_-_Mavail_diff_vs_age}
\end{center}
\end{figure*}
%- - - - - - - - - - - - - - - - - - - - - - - - - - - - - - - - - - - - - - - - - - - - - - - - - - - - - - - - - - - - - - - - - - - - - - - - - - - - - - - - - - - - - - - - - - - - 
% Cont SFH = logt_L vs model age
%- - - - - - - - - - - - - - - - - - - - - - - - - - - - - - - - - - - - - - - - - - - - - - - - - - - - - - - - - - - - - - - - - - - - - - - - - - - - - - - - - - - - - - - - - - - - 
\begin{figure*}
\begin{center}
\includegraphics[width=0.95\textwidth]{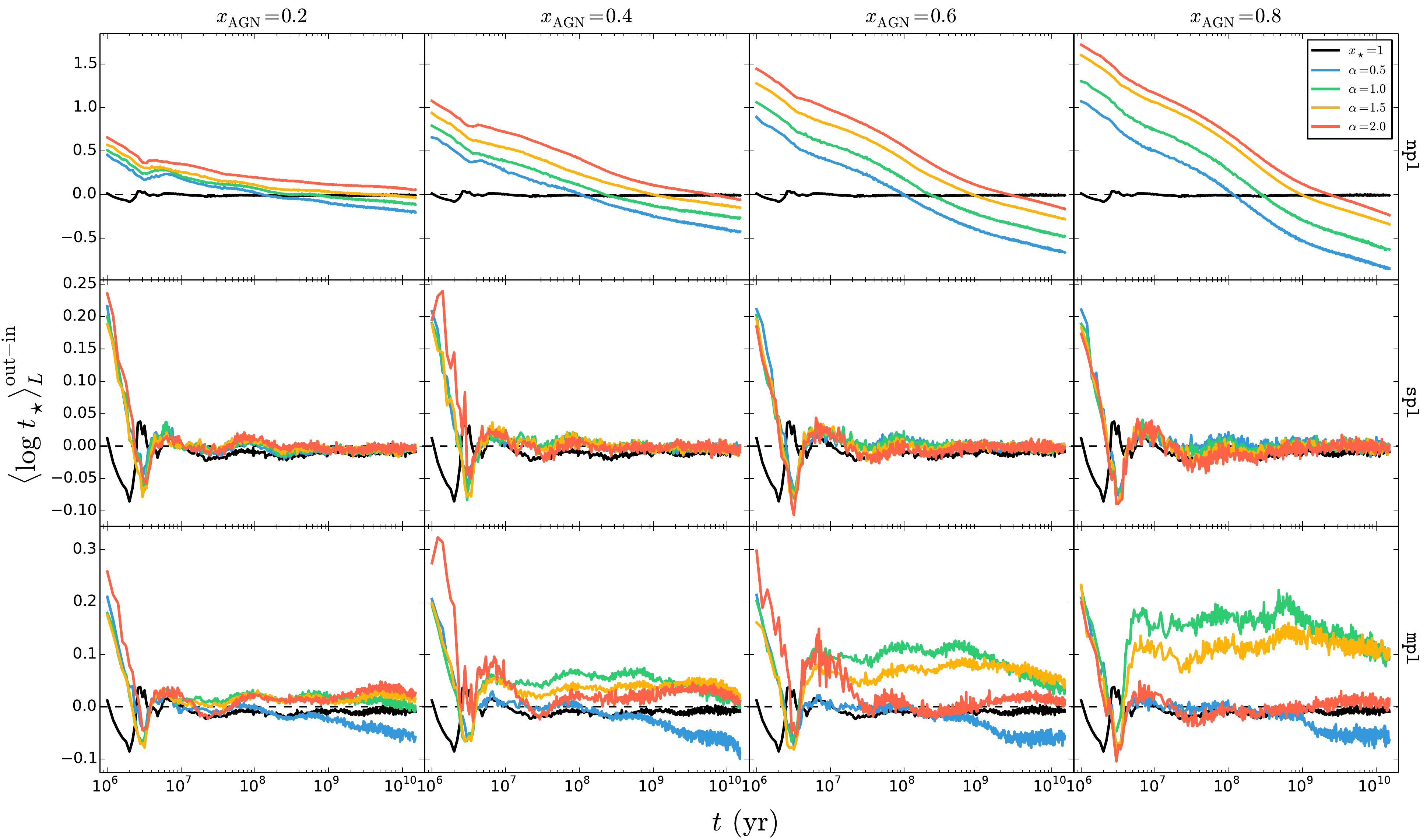}
\caption{Difference in the light-weighted mean stellar age $ \langle \log t_{\star} \rangle_L$ between PSS and ESS values as a function of model age $t$ for a continuous SFH. Panel configuration and legend details are analogous to those of Fig. \ref{Fig:Tau1_-_Mavail_diff_vs_age}.}
\label{Fig:Cont_-_logtL_diff_vs_age}
\end{center}
\end{figure*}
%- - - - - - - - - - - - - - - - - - - - - - - - - - - - - - - - - - - - - - - - - - - - - - - - - - - - - - - - - - - - - - - - - - - - - - - - - - - - - - - - - - - - - - - - - - - - 
% Cont SFH = logt_M vs model age
%- - - - - - - - - - - - - - - - - - - - - - - - - - - - - - - - - - - - - - - - - - - - - - - - - - - - - - - - - - - - - - - - - - - - - - - - - - - - - - - - - - - - - - - - - - - - 
\begin{figure*}
\begin{center}
\includegraphics[width=0.95\textwidth]{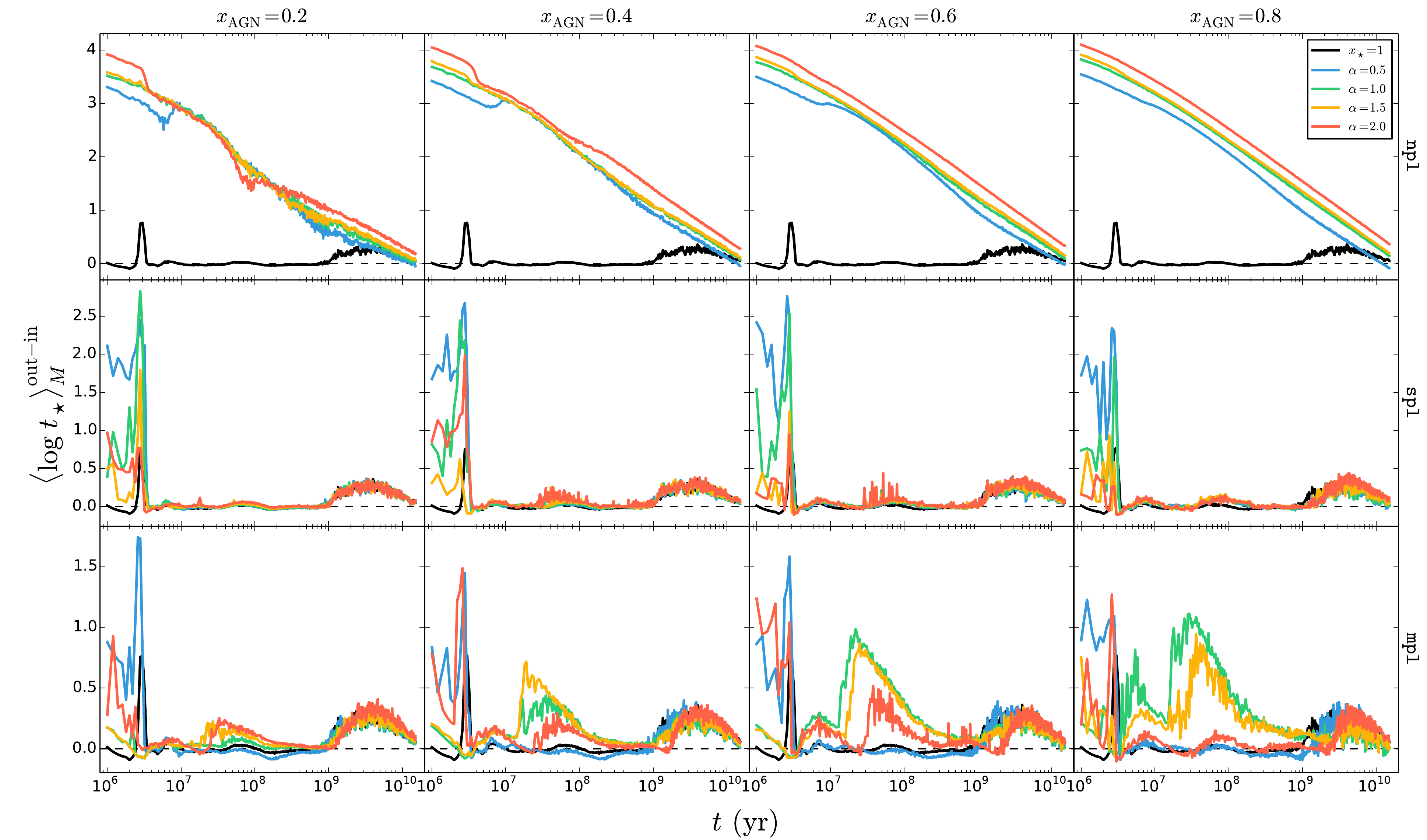}
\caption{Difference in the mass-weighted mean stellar age $\langle \log t_{\star} \rangle_M$ between PSS and ESS values as a function of model age $t$ for a continuous SFH. Panel configuration and legend details are analogous to those of Fig. \ref{Fig:Tau1_-_Mavail_diff_vs_age}. }
\label{Fig:Cont_-_logtM_diff_vs_age}
\end{center}
\end{figure*}
%- - - - - - - - - - - - - - - - - - - - - - - - - - - - - - - - - - - - - - - - - - - - - - - - - - - - - - - - - - - - - - - - - - - - - - - - - - - - - - - - - - - - - - - - - - - - 
% Cont SFH = logZ_L vs model age        
%- - - - - - - - - - - - - - - - - - - - - - - - - - - - - - - - - - - - - - - - - - - - - - - - - - - - - - - - - - - - - - - - - - - - - - - - - - - - - - - - - - - - - - - - - - - - 
\begin{figure*}
\begin{center}
\includegraphics[width=0.95\textwidth]{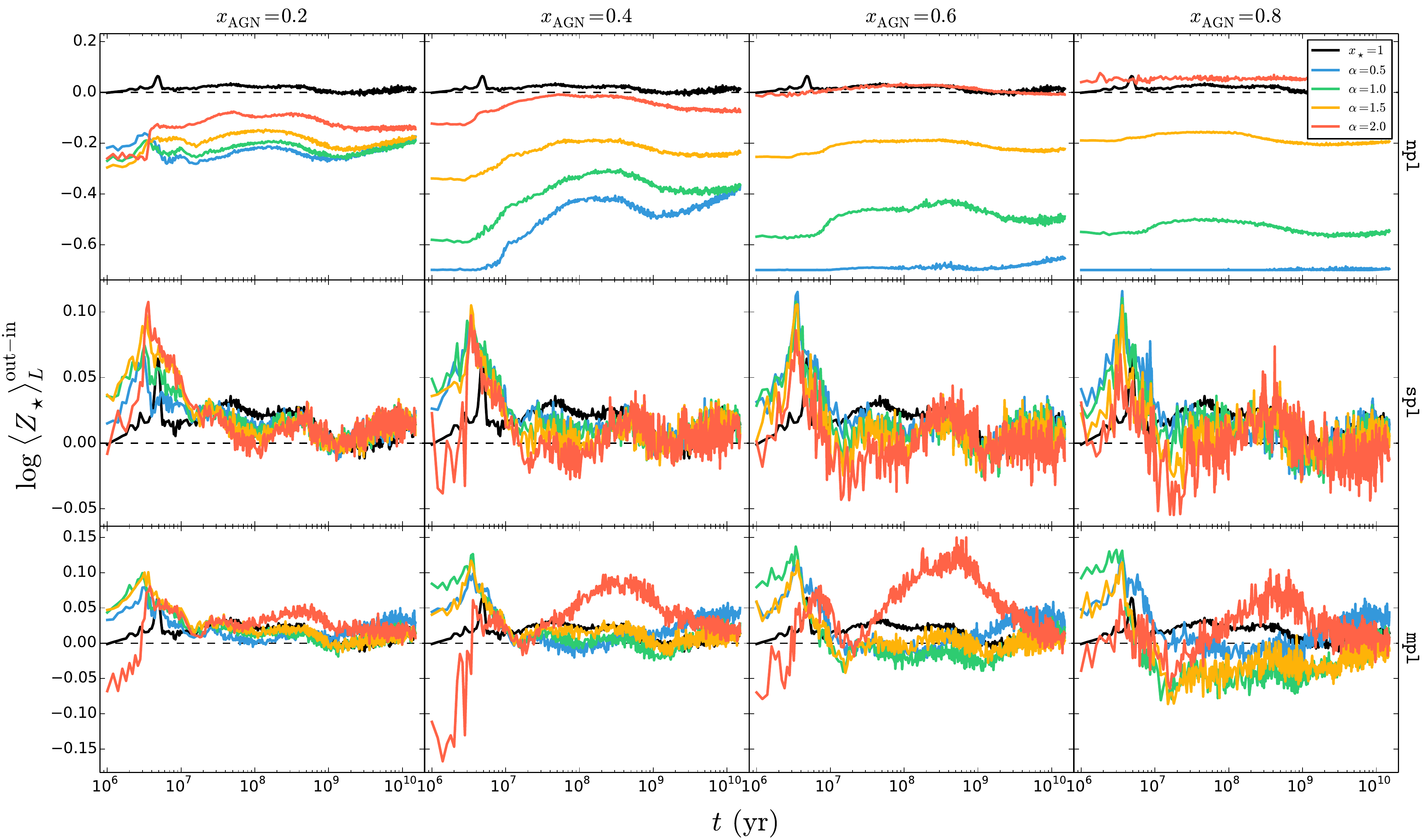}
\caption{Difference in the light-weighted mean stellar metallicity $\log \langle Z_{\star} \rangle_L$ between PSS and ESS values as a function of model age $t$ for a continuous SFH. Panel configuration and legend details are analogous to those of Fig. \ref{Fig:Tau1_-_Mavail_diff_vs_age}.}
\label{Fig:Cont_-_logZL_diff_vs_age}
\end{center}
\end{figure*}
%- - - - - - - - - - - - - - - - - - - - - - - - - - - - - - - - - - - - - - - - - - - - - - - - - - - - - - - - - - - - - - - - - - - - - - - - - - - - - - - - - - - - - - - - - - - - 
% Cont SFH = logZ_M vs model age        
%- - - - - - - - - - - - - - - - - - - - - - - - - - - - - - - - - - - - - - - - - - - - - - - - - - - - - - - - - - - - - - - - - - - - - - - - - - - - - - - - - - - - - - - - - - - - 
\begin{figure*}
\begin{center}
\includegraphics[width=0.95\textwidth]{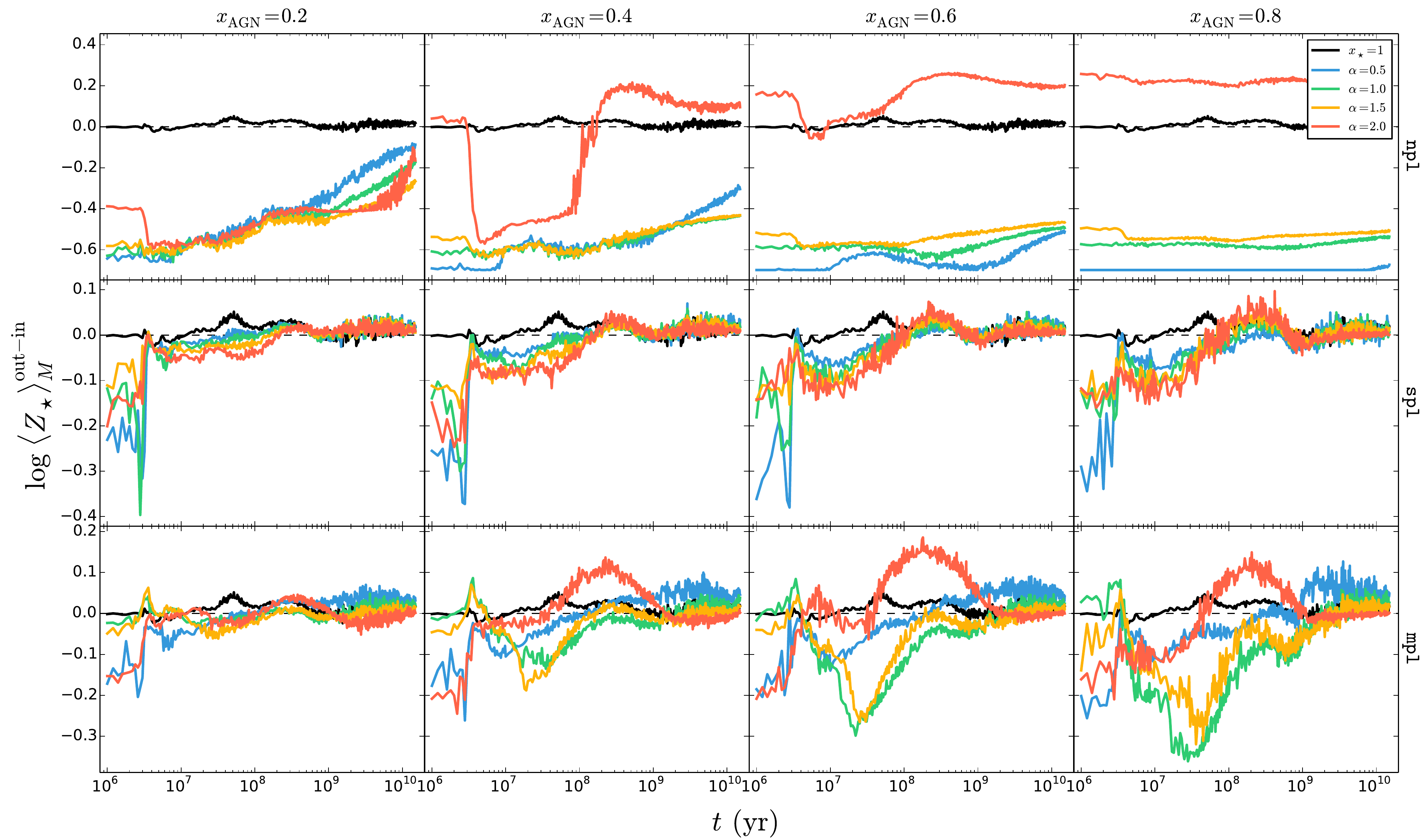}
\caption{Difference in the mass-weighted mean stellar metallicity difference $\log \langle Z_{\star} \rangle_M$ PSS and ESS values as a function of model age $t$ for a continuous SFH. Panel configuration and legend details are analogous to those of Fig. \ref{Fig:Tau1_-_Mavail_diff_vs_age}. }
\label{Fig:Cont_-_logZM_diff_vs_age}
\end{center}
\end{figure*}
%- - - - - - - - - - - - - - - - - - - - - - - - - - - - - - - - - - - - - - - - - - - - - - - - - - - - - - - - - - - - - - - - - - - - - - - - - - - - - - - - - - - - - - - - - - - - 
% Cont SFH = Delta_x_AGN vs model age
%- - - - - - - - - - - - - - - - - - - - - - - - - - - - - - - - - - - - - - - - - - - - - - - - - - - - - - - - - - - - - - - - - - - - - - - - - - - - - - - - - - - - - - - - - - - - 
%- - - - - - - - - - - - - - - - - - - - - - - - - - - - - - - - - - - - - - - - - - - - - - - - - - - - - - - - - - - - - - - - - - - - - - - - - - - - - - - - - - - - - - - - - - - - 

%- - - - - - - - - - - - - - - - - - - - - - - - - - - - - - - - - - - - - - - - - - - - - - - - - - - - - - - - - - - - - - - - - - - - - - - - - - - - - - - - - - - - - - - - - - - - - 
%- - - - - - - - - - - - - - - - - - - - - - - - - - - - - - - - - - - - - - - - - - - - - - - - - - - - - - - - - - - - - - - - - - - - - - - - - - - - - - - - - - - - - - - - - - - - - 
\label{lastpage}


\begin{thebibliography}{}
\small

% -- A --
\bibitem[\protect\citeauthoryear{Anders et al.}{2004}]{Anders_etal_2004} Anders, P., Bissantz, N., Fritze-v. Alvensleben, U., de Grijs, R. 2004, MNRAS, 1, 196
\bibitem[\protect\citeauthoryear{Asari et al.}{2007}]{Asari_etal_2007} {Asari}, N.~V., {Cid Fernandes}, R., {Stasi{\'n}ska}, G., {Torres-Papaqui}, J.~P., {Mateus}, A., {Sodr{\'e}}, L., {Schoenell}, W., {Gomes}, J.~M. 2007, MNRAS, 381, 263
\bibitem[Antonucci(1993)]{Antonucci1993}Antonucci R., 1993, ARA\&A, 31, 473
% -- B --
\bibitem[\protect\citeauthoryear{Baade}{1944a}]{Baade_1944a} Baade, W. 1944 ApJ, 100 ,137
\bibitem[\protect\citeauthoryear{Baade}{1944b}]{Baade_1944b} Baade, W. 1944 ApJ, 100 ,147
\bibitem[\protect\citeauthoryear{Ben\' itez et al.}{2013}]{Benitez_etal_2013} Ben\' itez, E., M\' endez-Abreu, J., Fuentes-Carrera, I., Cruz-Gonz\' alez, I., Mart\' inez, B., L\' opez-Martin, L., Jim\' enez-Bail\' on, E., Le\' on-Tavares, J., Chavushyan, V.~H. 2013, ApJ, 763, 36
\bibitem[\protect\citeauthoryear{Bica}{1988}]{Bica_1988} Bica E. 1988, A\&A, 195, 76
\bibitem[\protect\citeauthoryear{Bon, Popovi\'c \& Bon}{2014}]{Bon_Popovic_Bon_2014}Bon, N., Popovi\'c, L. C., Bon, E 2014, AdSpR, 54, 1389B
\bibitem[\protect\citeauthoryear{Boisson et al.}{2000}]{Boisson_etal_2000}Boisson, C., Joly, M., Moultaka, J., Pelat, D.,  Serote Roos, M. 2000, A\&A, 357, 850B
\bibitem[\protect\citeauthoryear{Bressan, Chiosi \& Tantalo}{1996}]{Bressan_Chiosi_Tantalo_1996} Bressan, A., Chiosi, C., Tantalo, R. 1996, ASPC, 98, 49
\bibitem[\protect\citeauthoryear{Bruzual \& Charlot}{2003}]{Bruzual_Charlot_2003} Bruzual, G., Charlot, S. 2003, MNRAS, 344, 1000
% -- C --
\bibitem[\protect\citeauthoryear{Calzetti et al.}{2000}]{Calzetti_etal_2000} Calzetti, D., Armus, L., Bohlin, R.~C., Kinney, A.~L., Koornneef, K., Storchi-Bergmann, T. 2000, ApJ, 533, 682
\bibitem[\protect\citeauthoryear{Cardoso, Gomes \& Papaderos}{2016}]{Cardoso_Gomes_Papaderos_2016} {Cardoso}, L.~S.~M.,  {Gomes}, J.-M., {Papaderos}, P. 2016, A\&A, 594, L2
\bibitem[\protect\citeauthoryear{Cervi\~no}{2013}]{Cervino_2013} Cervi$\tilde{\mathrm{n}}$o, M. 2013, NewAR, 57, 123
\bibitem[\protect\citeauthoryear{Chabrier}{2003}]{Chabrier_2003} Chabrier, G. 2003, PASP, 115, 763
\bibitem[\protect\citeauthoryear{Ciesla et al.}{2015}]{Ciesla_etal_2015} Ciesla, L., Charmandaris, V., Georgakakis, A., Bernhard, E., Mitchell, P.~D., Buat, V., Elbaz, D., LeFloc'h, E., Lacey, C.~G., Magdis, G.~E., Xilouris, M. 2015, A\&A, 576, 10
\bibitem[\protect\citeauthoryear{da Cunha, Charlot \& Elbaz}{2008}]{daCunha_etal_2008} da Cunha E., Charlot S., Elbaz D. 2008, MNRAS, 388, 1595
\bibitem[\protect\citeauthoryear{Cid Fernandes \& Terlevich}{1995}]{Cid1995}Cid Fernandes R., Terlevich R. 1995, MNRAS 272, 423
\bibitem[\protect\citeauthoryear{Cid Fernandes, Storchi-Bergmann \& Schmitt}{1998}]{CidFernandes1998}Cid Fernandes, R. Jr., Storchi-Bergmann, T., Schmitt, H. R. 1998, MNRAS, 297, 579C
\bibitem[\protect\citeauthoryear{Cid Fernandes et al.}{2004}]{CidFernandes_etal_2004} Cid Fernandes, R., Gu, Q., Melnick,
  J., Terlevich, E., Terlevich, R., Kunth, D., Rodrigues Lacerda, R., Joguet,
  B. 2004, MNRAS, 355, 273
\bibitem[\protect\citeauthoryear{Cid Fernandes et al. }{2005}]{CidFernandes_etal_2005} Cid Fernandes, R., Mateus, A., Sodr\' e, L., Stasi\' nska, G., Gomes, J.~M. 2005, MNRAS, 358, 363
\bibitem[\protect\citeauthoryear{Cid Fernandes et al. }{2010}]{CidFernandes_etal_2010} Cid Fernandes, R., Stasi\' nska, G., Schlickmann, M.~S., Mateus, A., Vale Asari, N., Schoenell, W., Sodr\' e, L. 2010, MNRAS, 403, 1036
\bibitem[\protect\citeauthoryear{Cid Fernandes et al. }{2011}]{CidFernandes_etal_2011} Cid Fernandes, R., Stasi\' nska, G., Mateus, A., Vale Asari, N. 2011, MNRAS, 413, 1687
\bibitem[\protect\citeauthoryear{Cid Fernandes et al. }{2014}]{CidFernandes_etal_2014} Cid Fernandes, R., Gonz\' alez Delgado, R.~M., Garc\' ia Benito, R., P\' erez, E., de Amorim, A.~L., S\' anchez, S.~F., Husemann, B., Falc\' on Barroso, J., L\' opez-Fern\' andez, R., S\' anchez-Bl\' azquez, P., Vale Asari, N., Vazdekis, A., Walcher, C.~J., Mast, D. 2014, A\&A, 561, 19
\bibitem[\protect\citeauthoryear{Coelho et al.}{2007}]{Coelho_etal_2007} Coelho, P., Bruzual, G., Charlot, S., Weiss, A., Barbuy, B., Ferguson, J. W. 2007, MNRAS, 12, 498
\bibitem[\protect\citeauthoryear{Conroy}{2013}]{Conroy_2013} Conroy, C. 2013, ARA\&A, 51, 393
% -- D --
\bibitem[\protect\citeauthoryear{Dressler}{1984}]{Dressler1984}Dressler, A. 1984, ApJ, 286, 97D
% -- E --
\bibitem[\protect\citeauthoryear{Eracleous \& Halpern}{2001}]{Eracleous_Halpern_2001} Eracleous, M., Halpern, J.~P. 2001, ApJ, 554, 240
% -- F --
\bibitem[\protect\citeauthoryear{Faber}{1972}]{Faber_1972} Faber, S. M. 1972,
  A\&A, 20, 361
\bibitem[\protect\citeauthoryear{Ferland \& Netzer}{1983}]{Ferland_Netzer_1983} Ferland, G.~J., Netzer, H. 1983, ApJ, 264, 105
\bibitem[\protect\citeauthoryear{Fioc \& Rocca-Volmerange}{1997}]{Fioc_RoccaVolmerange_1997} Fioc, M., Rocca-Volmerange, B. 1997, A\&A, 10, 950
% -- G --
\bibitem[\protect\citeauthoryear{Gavazzi et al.}{2002}]{Gavazzi_etal_2002} Gavazzi, G., Bonfanti, C., Sanvito, G., Boselli, A., Scodeggio, M. 2002 ApJ, 576, 135  
\bibitem[\protect\citeauthoryear{Garcia-Rissmann et al.}{2005}]{GarciaRissmann_etal_2005} Garcia-Rissmann, A., Vega, L. R., Asari, N.~V., Cid Fernandes, R., Schmitt, H., Gonz\'alez Delgado, R.~M., Storchi-Bergmann, T. 2005, MNRAS, 359, 765
\bibitem[\protect\citeauthoryear{Goerdt \& Kollatschny}{1998}]{Goerdt_Kollatschny_1998} Goerdt, A., Kollatschny, W. 1998, A\&A, 337, 699
\bibitem[\protect\citeauthoryear{Gomes et al.}{2016}]{Gomes_etal_2016c}  {Gomes}, J.~M., {Papaderos}, P., {Kehrig}, C., {V{\'{\i}}lchez}, J.~M.,  {Lehnert}, M.~D., {S{\'a}nchez}, S.~F., {Ziegler}, B.,  {Breda}, I., {Dos Reis}, S.~N., {Iglesias-P{\'a}ramo}, J.,  {Bland-Hawthorn}, J., {Galbany}, L., {Bomans}, D.~J.,  {Rosales-Ortega}, F.~F., {Cid Fernandes}, R., {Walcher}, C.~J.,  {Falc{\'o}n-Barroso}, J., {Garc{\'{\i}}a-Benito}, R.,  {M{\'a}rquez}, I., {Del Olmo}, A., {Masegosa}, J.,  {Moll{\'a}}, M., {Marino}, R.~A., {Gonz{\'a}lez Delgado}, R.~M.,  {L{\'o}pez-S{\'a}nchez}, {\'A}.~R.,  CALIFA Collaboration 2016, A\&A, 588, 68
% -- H --
\bibitem[\protect\citeauthoryear{Hayward \& Smith}{2015}]{Hayward_Smith_2015} Hayward, C.~C.; Smith, D.~J.~B. 2015, MNRAS, 446, 1512
\bibitem[\protect\citeauthoryear{Heavens, Jimenez \& Lahav}{2000}]{Heavens_Jimenez_Lahav_2000} Heavens, A.~F., Jimenez, R., Lahav, O. 2000, MNRAS, 317, 965
\bibitem[\protect\citeauthoryear{Heavens et al.}{2004}]{Heavens_etal_2004} Heavens, A., Panter, B., Jimenez, R., Dunlop, J. 2004, Nat, 428, 625
\bibitem[\protect\citeauthoryear{Heckman}{1980}]{Heckman_1980} Heckman, T.~M. 1980, A\&A, 88, 311
\bibitem[Heckman et al.(1995)]{Heckman1995}Heckman T.M., Krolik J., Meurer G.,  et al., 1995 ApJ 452, 549
\bibitem[\protect\citeauthoryear{Ho, Filippenko \& Sargent}{1995}]{Ho_Filippenko_Sargent_1995}  Ho, L. C., Filippenko, A. V., Sargent, W. L. 1995, ApJS, 98, 477
\bibitem[\protect\citeauthoryear{Ho, Filippenko \& Sargent}{1997}]{Ho_Filippenko_Sargent_1997}  Ho, L. C., Filippenko, A. V., Sargent, W. L. W. 1997, ApJS, 112, 315
\bibitem[\protect\citeauthoryear{Ho, Filippenko \& Sargent}{2003}]{Ho_Filippenko_Sargent_2003}  Ho, L. C., Filippenko, A. V., Sargent, W. L. W. 2003, ApJ, 583, 159
\bibitem[\protect\citeauthoryear{Ho}{2008}]{Ho_2008}Ho, L. 2008, ARA\&A, 46, 475H
\bibitem[\protect\citeauthoryear{Ho \& Kim}{2009}]{Ho_Kim_2009} Ho, L.~C., Kim, M. 2009, ApJS, 184, 398
% -- K -- 
\bibitem[\protect\citeauthoryear{Kauffmann et al.}{2003}]{Kauffmann_etal_2003c} Kauffmann, G., Heckman, T. M., Tremonti, C., Brinchmann, J., Charlot, S., White, S. D. M., Ridgway, S. E., Brinkmann, J., Fukugita, M., Hall, P. B., Ivezi\' c, \v Z., Richards, G.~T., Schneider, D. P. 2003, MNRAS, 346, 1055
\bibitem[\protect\citeauthoryear{Kehrig et al.}{2012}]{Kehrig_etal_2012} Kehrig, C., Monreal-Ibero, A., Papaderos, P., V\' ilchez, J.~M., Gomes, J.~M., Masegosa, J., S\' anchez, S.~F., Lehnert, M.~D., Cid Fernandes, R., Bland-Hawthorn, J., Bomans, D.~J., Marquez, I., Mast, D., Aguerri, J.~A.~L., L\' opez-S\' anchez, Á.~R., Marino, R.~A., Pasquali, A., Perez, I., Roth, M.~M., S\' anchez-Bl\' azquez, P., Ziegler, B. 2012, A\&A, 540, 11
\bibitem[\protect\citeauthoryear{Koleva et al.}{2009}]{Koleva_etal_2009} Koleva, M., Prugniel, Ph., Bouchard, A., Wu, Y. 2009, A\&A, 7, 1269
\bibitem[\protect\citeauthoryear{Koski}{1978}]{Koski_1978} Koski, A. T. 1978, ApJ, .223, 56
\bibitem[\protect\citeauthoryear{Kotulla et al.}{2009}]{Kotulla_etal_2009} Kotulla, R., Fritze, U., Weilbacher, P., Anders, P. 2009, MNRAS, 6, 462
\bibitem[\protect\citeauthoryear{Kr\"uger, Fritze-v. Alvensleben \& Loose}{1995}]{Kruger_etal_1995} Kr\"uger, H., Fritze-v. Alvensleben, U., Loose, H.-H. 1995, A\&A, 11, 41
% -- L -- 
\bibitem[\protect\citeauthoryear{Leitherer et al.}{1999}]{Leitherer_etal_1999} Leitherer, C., Schaerer, D., Goldader, J.~D., Gonz{\'a}lez Delgado, R.~M., Robert, C., Kune, D.~F., de Mello, D.~F., Devost, D., Heckman, T.~M. 1999, ApJS, 7, 3
\bibitem[\protect\citeauthoryear{Le Borgne et al.}{2003}]{LeBorgne_etal_2003} Le Borgne, J.-F., Bruzual, G., Pell{\'o}, R., Lan{\c c}on, A., Rocca-Volmerange, B., Sanahuja, B., Schaerer, D., Soubiran, C., V{\'{\i}}lchez-G{\'o}mez, R. 2003, A\&A, 402, 433
\bibitem[\protect\citeauthoryear{Le Borgne et al.}{2004}]{LeBorgne_etal_2004} Le Borgne, D., Rocca-Volmerange, B., Prugniel, P., Lan{\c c}on, A., Fioc, M., Soubiran, C. 2004, A\&A, 10, 881
% -- M -- 
\bibitem[\protect\citeauthoryear{MacArthur et al.}{2009}]{MarcArthur_etal_2009} MacArthur, L. A., Gonz{\'a}lez, J. J., Courteau, S. 2009, MNRAS, 5, 28
\bibitem[\protect\citeauthoryear{Maraston}{2005}]{Maraston_2005} Maraston, C. 2005, MNRAS, 362, 799
\bibitem[\protect\citeauthoryear{Mateus et al.}{2006}]{Mateus_etal_2006} Mateus, A., Sodr\' e, L., Cid Fernandes, R., Stasi\' nska, G., Schoenell, W., Gomes, J. M. 2006, MNRAS, 370, 721
\bibitem[\protect\citeauthoryear{Mathews \& Ferland}{1987}]{Mathews_Ferland_1987} Mathews, W. G., Ferland, G. J. 1987, ApJ, 323, 456
\bibitem[\protect\citeauthoryear{Moll{\'a} et al.}{2009}]{Molla_etal_2009} Moll{\'a}, M., Garc{\'\i}a-Vargas, M. L., Bressan, A. 2009, MNRAS, 9, 451
\bibitem[\protect\citeauthoryear{Morgan}{1956}]{Morgan_1956} Morgan W. W. 1956, PASP, 68, 509
\bibitem[\protect\citeauthoryear{Moultaka \& Pelat}{2000}]{Moultaka_Pelat_2000} Moultaka, J., Pelat, D. 2000, MNRAS, 314, 409
\bibitem[\protect\citeauthoryear{Moultaka et al.}{2004}]{Moultaka_etal_2004} Moultaka, J., Boisson, C., Joly, M., Pelat, D. 2004, A\&A, 420, 459
\bibitem[\protect\citeauthoryear{Moultaka}{2005}]{Moultaka_2005} Moultaka, J. 2005, A\&A, 430, 95
% -- N --
\bibitem[Netzer(2015)]{Netzer2015}Netzer, H. 2015, ARA\&A, 53, 365N
\bibitem[Nelson \& Whittle(1995)]{NelsonWhittle1995}Nelson, C. H., N., Whittle, M. 1995, ApJS, 99, 67N
\bibitem[\protect\citeauthoryear{Noll et al.}{2009}]{Noll_etal_2009} Noll, S., Burgarella, D., Giovannoli, E., Buat, V., Marcillac, D., Mu$\tilde{\mathrm{n}}$oz-Mateos, J. C. 2009, A\&A, 507, 1793
% -- O --
\bibitem[\protect\citeauthoryear{O'Connell}{1976}]{OConnell_1976} O'Connell, R. W. 1976, ApJ, 206, 370
\bibitem[\protect\citeauthoryear{O'Connell}{1980}]{OConnell_1980} O'Connell, R. W. 1980, ApJ, 236, 430
\bibitem[\protect\citeauthoryear{Ocvirk et al.}{2006b}]{Ocvirk_etal_2006b} Ocvirk, P., Pichon, C., Lan{\c c}on, A., Thi{\'e}baut, E. 2006a, MNRAS, 365, 46
\bibitem[\protect\citeauthoryear{Ocvirk et al.}{2006a}]{Ocvirk_etal_2006a} Ocvirk, P., Pichon, C., Lan{\c c}on, A., Thi{\'e}baut, E. 2006b, MNRAS, 365, 74
\bibitem[\protect\citeauthoryear{Oke, Neugebauer \& Becklin}{1970}]{Oke_Neugebauer_Becklin_1970} Oke, J. B., Neugebauer, G., Becklin, E. E. 1970, ApJ, 159, 341
% -- P --
\bibitem[\protect\citeauthoryear{Papaderos et al.}{2013}]{Papaderos_etal_2013} Papaderos, P., Gomes, J. M., V\' ilchez, J. M., Kehrig, C., Lehnert, M. D., Ziegler, B., S\' anchez, S. F., Husemann, B., Monreal-Ibero, A., Garc\'ia-Benito, R., Bland-Hawthorn, J., Cortijo-Ferrero, C., de Lorenzo-C\'aceres, A., del Olmo, A., Falc\' on-Barroso, J., Galbany, L., Iglesias-P\' aramo, J., L\' opez S\' anchez, \' A. R., Marquez, I., Moll\' a, M., Mast, D., van de Ven, G., Wisotzki, L. 2013, A\&A, 555, L1
\bibitem[\protect\citeauthoryear{Pelat}{1997}]{Pelat_1997} Pelat, D. 1997, MNRAS, 284, 365
\bibitem[\protect\citeauthoryear{Pelat}{1998}]{Pelat_1998} Pelat, D. 1998, MNRAS, 299, 877
% -- Q --
% -- R --
\bibitem[\protect\citeauthoryear{Renzini}{1981}]{Renzini_1981} Renzini, A. 1981. Ann. Phys. 6, 87
\bibitem[\protect\citeauthoryear{Renzini \& Buzzoni}{1986}]{Renzini_Buzzoni_1986} Renzini, A., Buzzoni, A. 1986, ASSL, 122, 195
\bibitem[\protect\citeauthoryear{Ribeiro et al.}{2016}]{Ribeiro_etal_2016} Ribeiro, B., Lobo, C., Ant{\'o}n, S.,  Gomes, J.~M., \& Papaderos, P. 2016, MNRAS, 456, 3899
\bibitem[\protect\citeauthoryear{Rocca-Volmerange \& Guiderdoni}{1988}]{RoccaVolmerange_Guiderdoni_1988} Rocca-Volmerange, B., Guiderdoni, B. 1988, A\&ASS, 75, 91
\bibitem[\protect\citeauthoryear{Roche et al.}{2016}]{Roche_etal_2016} Roche, N., Humphrey, A., Lagos, P., Papaderos, P., Silva, M., Cardoso, L.~S.~M., Gomes, J.~M. 2016, MNRAS, 459, 4259
% -- S --
\bibitem[\protect\citeauthoryear{Schmitt, Storchi-Bergmann \& Cid Fernandes}{1999}]{Schmitt_StorchiBergmann_CidFernandes_1999} Schmitt, H. R., Storchi-Bergmann, T., Cid Fernandes, R. 1999, MNRAS, 303, 173
\bibitem[\protect\citeauthoryear{Seyfert}{1943}]{Seyfert_1943} Seyfert, C. K. 1943, ApJ, 97, 28
\bibitem[\protect\citeauthoryear{Serote Roos et al.}{1998}]{SeroteRoos_etal_1998} Serote Roos, M., Boisson, C., Joly, M., Ward, M. J. 1998, MNRAS, 301, 1
\bibitem[\protect\citeauthoryear{Stasi\' nska}{1984a}]{Stasinska_1984a} Stasi\' nska, G. 1984a, A\&AS, 55, 15
\bibitem[\protect\citeauthoryear{Stasi\' nska}{1984b}]{Stasinska_1984b} Stasi\' nska, G. 1984b, A\&A, 135, 341
\bibitem[\protect\citeauthoryear{Stasi\' nska et al.}{2008}]{Stasinska_etal_2008} Stasi\' nska, G., Vale Asari, N., Cid Fernandes, R., Gomes, J. M., Schlickmann, M., Mateus, A., Schoenell, W., Sodr\' e, L., Jr. 2008, MNRAS, 391, 29
\bibitem[\protect\citeauthoryear{Stasi\' nska et al.}{2015}]{Stasinska_etal_2015} Stasi\' nska, G., Costa-Duarte, M. V., Vale Asari, N., Cid Fernandes, R., Sodr\' e, L. 2015 MNRAS, 449, 559
\bibitem[\protect\citeauthoryear{Storchi-Bergmann, Baldwin \& Wilson}{1993}]{StochiBergmann_etal_1993}Storchi-Bergmann, T., Baldwin, J. A., Wilson, A. S. 1993, ApJ, 410L, 11S
\bibitem[\protect\citeauthoryear{Storchi-Bergmann, Cid Fernandes \& Schmitt}{1998}]{StochiBergmann_etal_1998}Storchi-Bergmann, T., Cid Fernandes, R., Schmitt, H. R. 1998, ApJ, 501, 94S
\bibitem[\protect\citeauthoryear{Spinrad \& Taylor}{1972}]{Spinrad_Taylor_1972} Spinrad, H., Taylor, B. J. 1972, ApJ, 171, 397
% -- T --
\bibitem[Tran(1995)]{Tran1995}Tran, H.~D. 1995, ApJ, 440, 597T
\bibitem[\protect\citeauthoryear{Terlevich, D\'iaz \& Terlevich}{1990}]{Terlevich_Diaz_Terlevich_1990} Terlevich, E., D\'iaz, A.~I., Terlevich, R. 1990, MNRAS, 242, 271
\bibitem[\protect\citeauthoryear{Tinsley}{1968}]{Tinsley_1968} Tinsley, B. M. 1968, ApJ, 151, 547
\bibitem[\protect\citeauthoryear{Tinsley}{1980}]{Tinsley_1980} Tinsley, B. M. 1980, FCPh, 5, 287
\bibitem[\protect\citeauthoryear{Tojeiro et al.}{2007}]{Tojeiro_etal_2007} Tojeiro, R., Heavens, A.~F., Jimenez, R., Panter, B. 2007 MNRAS, 11, 1252
% -- V --
\bibitem[\protect\citeauthoryear{Vazdekis et al.}{2010}]{Vazdekis_etal_2010}  Vazdekis, A., S{\'a}nchez-Bl{\'a}zquez, P., Falc{\'o}n-Barroso, J., Cenarro, A. J., Beasley, M.~A., Cardiel, N., Gorgas, J., Peletier, R.~F. 2010, MNRAS, 6, 1639
\bibitem[\protect\citeauthoryear{Vega et al.}{2009}]{Vega_etal_2009} Vega, L.~R., Asari, N.~V., Cid Fernandes, R., Garcia-Rissmann, A., Storchi-Bergmann, T., González Delgado, R.~M., Schmitt, H. 2009, MNRAS, 393, 846
\bibitem[\protect\citeauthoryear{Veilleux \& Osterbrock}{1987}]{Veilleux_Osterbrock_1987} Veilleux, S., Osterbrock, D.~E. 1987, ApJS, 63, 295
% -- W --
\bibitem[\protect\citeauthoryear{Walcher et al.}{2011}]{Walcher_etal_2011} Walcher, J., Groves, B., Budav\' ari, T; Dale, D. 2011, Ap\&SS, 331, 1
\bibitem[\protect\citeauthoryear{Whipple}{1935}]{Whipple_1935} Whipple, Fred L. 1935, Harvard College Observatory Circular, 404, 1
\bibitem[\protect\citeauthoryear{Wood}{1966}]{Wood_1966} Wood D. B. 1966, ApJ, 145, 36 
\bibitem[\protect\citeauthoryear{Worthey}{1994}]{Worthey_1994} Worthey G. 1994, ApJS, 95, 107
% -- Y --
\bibitem[\protect\citeauthoryear{York et al.}{2000}]{York_etal_2000} {York}, D.~G., {Adelman}, J., {Anderson}, Jr., J.~E., {SDSS Collaboration} 2000 ApJ, 120, 1579
% -- Z --
\bibitem[\protect\citeauthoryear{Zackrisson et al.}{2001}]{Zackrisson_etal_2001} Zackrisson, E., Bergvall, N., Olofsson, K., Siebert, A. 2001, A\&A, 9, 814
\end{thebibliography}
\end{document}